\documentclass[preprint,twocolumn,5p]{elsarticle}
\usepackage{amsmath,amssymb,amsfonts}
\usepackage{subfigure}
\usepackage{hyperref}
\usepackage{url}
\usepackage{natbib}
\usepackage{multirow}
\usepackage{lineno}
\modulolinenumbers[5]

\begin{document}

\begin{frontmatter}
\title{Dual-embedding based Neural Collaborative Filtering for Recommender Systems}

% \author{
% 	\IEEEauthorblockN{Gongshan He}
% 	\IEEEauthorblockA{
% 		\textit{School of Computer Science} \\
% 		\textit{Wuhan University} \\
% 		hegongshan@whu.edu.cn
% 	}
% 	\and
% 	\IEEEauthorblockN{Dongxing Zhao}
% 	\IEEEauthorblockA{
% 		\textit{School of Computer Science} \\
% 		\textit{Wuhan University} \\
% 		dxchiu@whu.edu.cn
% 	}
% 	\and
% 	\IEEEauthorblockN{
% 		Lixing Ding* \thanks{* Corresponding author}
% 	}
% 	\IEEEauthorblockA{
% 		\textit{School of Computer Science} \\
% 		\textit{Wuhan University} \\
% 		Lxding@whu.edu.cn
% 	}
% }

\author{Gongshan He}
\ead{hegongshan@whu.edu.cn}

\author{Dongxing Zhao}
\ead{dxchiu@whu.edu.cn}

\author{Lixin Ding\corref{correspondingauthor}}
\cortext[correspondingauthor]{Corresponding author}
\ead{Lxding@whu.edu.cn}

\address{School of Computer Science, Wuhan University, Wuhan, Hubei Province 430072, PR China}

\begin{abstract}
Among various recommender techniques, collaborative filtering (CF) is the most successful one. And a key problem in CF is how to represent users and items. Previous works usually represent a user (an item) as a vector of latent factors (aka. \textit{embedding}) and then model the interactions between users and items based on the representations. Despite its effectiveness, we argue that it's insufficient to yield satisfactory embeddings for collaborative filtering. Inspired by the idea of SVD++ that represents users based on themselves and their interacted items, we propose a general collaborative filtering framework named DNCF, short for Dual-embedding based Neural Collaborative Filtering, to utilize historical interactions to enhance the representation. In addition to learning the primitive embedding for a user (an item), we introduce an additional embedding from the perspective of the interacted items (users) to augment the user (item) representation. Extensive experiments on four publicly datasets demonstrated the effectiveness of our proposed DNCF framework by comparing its performance with several traditional matrix factorization models and other state-of-the-art deep learning based recommender models.
\end{abstract}

\end{frontmatter}

\section{Introduction}
In the era of information explosion, users are often overwhelmed by numerous choices available online, which is named \textit{information overload}. Over the past decades, recommender systems have been intensively studied and extensively deployed in various scenarios, such as e-commerce and music platforms, to alleviate this problem. Collaborative Filtering (CF) \cite{DBLP:conf/www/SarwarKKR01,DBLP:journals/internet/LindenSY03} is one of the most successful recommender techniques and has been widely used to build personalized recommender systems, which utilizes collective wisdoms and experiences to generate recommendations.

The key challenge to design a CF model is: \textit{how to represent a user and an item} and \textit{how to model their interactions based on the representation} \cite{DBLP:conf/ijcai/0001DWTTC18}. As a dominant model in CF, Matrix Factorization (MF) \cite{DBLP:conf/nips/SalakhutdinovM07, DBLP:journals/computer/KorenBV09} characterizes users and items with latent vectors (aka. \textit{embedding}) in a shared latent space and then each user-item interaction is modeled as the inner product between the user embedding and item embedding. Many extensions have been developed for MF from both the modeling perspective \cite{paterek:improving, DBLP:conf/kdd/Koren08} and learning perspective \cite{DBLP:conf/uai/RendleFGS09}. For example, NSVD \cite{paterek:improving} characterizes users by the items that they have rated. Specifically, a user embedding is represented by combining the embeddings of all items rated by the user. 
% FISM \cite{DBLP:conf/kdd/KabburNK13} is a widely used item-based CF method, which learns the item-item similarity matrix as a product of two low-dimensional latent factor matrices. It can also be viewed as a matrix factorization method, which represents a user by the embeddings of all items interacted by the user.
To further step, SVD++ \cite{DBLP:conf/kdd/Koren08} represents a user by integrating the embeddings of the items interacted by the user with primitive user embedding.

In recent years, deep learning methods achieve tremendous success in many fields, such as computer vision \cite{DBLP:conf/cvpr/HeZRS16} and natural language processing \cite{DBLP:conf/naacl/DevlinCLT19}. There are also many works applying deep learning to recommender systems. Neural matrix factorization (NeuMF) \cite{DBLP:conf/www/HeLZNHC17} represents a user or an item with an ID and learns the interactions by fusing the linear MF and non-lineaer multi-layer perceptron (MLP) models. DeepMF \cite{DBLP:conf/ijcai/XueDZHC17} feeds related rating vectors into MLP to learn users' (items') embeddings and then uses cosine similarity as interaction function to predict the relevance score.

Inspired by NSVD and NeuMF, DELF \cite{DBLP:conf/ijcai/ChengSZH18} is proposed to represent users or items by their dual embeddings. To be more specific, in addition to model primitive embeddings, it obtains additional embeddings from the perspective of the interacted users or  items and learn the interactions between users and items from four aspects: user-to-user, item-to-item, user-to-item (ID) and user-to-item (historical interactions). However, it only employs dual embeddings to learn four kinds of interaction functions for each user-item pair and does not combine two types of embeddings to get better user's (item's) representation.

To tackle this problem, we propose a general dual embeddings based CF framework named DNCF, short for \textbf{D}ual-embedding based \textbf{N}eural \textbf{C}ollaborative \textbf{F}iltering, to combine the strengths of the two types of embeddings. Specifically, we use the items interacted by users to augment user representation and use the users once interacted with items to enrich item representation. And then we employ deep neural network architecture to model the user-item interactions.

The main contributions of this work are as follows:
\begin{enumerate}
    \item We propose to combine users' (items') dual embeddings into final users' (items')  representation, namely, integrating their primary embeddings with additional embeddings obtained from the perspective of historical interactions to get the final representation.
    \item We devise a novel framework named \textbf{D}ual-embedding based \textbf{N}eural \textbf{C}ollaborative \textbf{F}iltering (DNCF) which models the interactions between users and items based on their dual embeddings.
    \item We conduct extensive experiments on four real-world datasets to demonstrate the effectiveness of our proposed DNCF approaches.
\end{enumerate}

The remaining of this article is organized as follows: Section \ref{section:preliminary} introduces the preliminaries for top-N recommendation. Section \ref{section:related_work} introduces some related works. Section \ref{section:proposed_method} presents our proposed DNCF framework in detail. Section \ref{section:experiment} illustrates the experimental results on four public datasets. Finally, we conclude this work and point out future research directions in section \ref{section:conclusion_futere_work}.

\section{Preliminaries\label{section:preliminary}}

\subsection{Problem Statement}
Let M and N denote the total number of users and items in the systems, respectively. Following \cite{DBLP:conf/www/HeLZNHC17,DBLP:conf/ijcai/XueDZHC17,DBLP:conf/aaai/DengHWLY19,DBLP:journals/tois/ChenCCR19}, we construct the user-item interaction matrix $\mathbf{Y} \in \mathbb{R}^{M \times N}$ from users' implicit feedback as follows,

\begin{equation}
    y_{ui} = 
    \begin{cases}
        1, & \mathrm{if\ interaction\ (user}\ u, \mathrm{item}\ i)\ \mathrm{is\ observed} \\
        0, & \mathrm{otherwise}
    \end{cases}
\end{equation}

For implicit feedback, all observed interactions are considered as noisy positive instances which reflect users' preference to some extent. However, there are no negative instances. 
A simple solution is to treat all unobserved interactions (i.e. the value of $y_{ui}$ is equal to 0) as negative feedback. Nevertheless, not all unobserved interactions are true negative instances. To be specific, an unobserved interaction does not necessarily mean user $u$ does not like item $i$. As a matter of fact, user $u$ may have never seen item $i$ since there are too many items in a system. Another approach is to sample negative instances from unobserved interactions \cite{DBLP:conf/icdm/PanZCLLSY08,DBLP:conf/www/HeLZNHC17}. In this work, we choose the latter, i.e. randomly sample negative instances from unobserved interactions without replacement.

The problem of recommendation with implicit feedback is to estimate the scores of unobserved entries in $\mathbf{Y}$, which are used for ranking the items.
Model-based approaches \cite{DBLP:conf/nips/SalakhutdinovM07,DBLP:journals/computer/KorenBV09} generally assume that data can be generated by an underlying model which can be formulated as 
\begin{equation}
    \hat{y}_{ui} = f(u,i|\Theta)
\end{equation}
where $\hat{y}_{ui}$ denotes the predicted score of interaction $y_{ui}$, $\Theta$ denotes model parameters, and $f$ denotes the function that maps model parameters to the predicted score.

\subsection{Learning the Model}

Most of existing approaches generally estimate parameters $\Theta$ through optimizing an objective function. Three types of objective functions are most commonly used in recommender systems —— point-wise loss \cite{DBLP:conf/icdm/HuKV08,DBLP:journals/tois/XueHWXLH19,DBLP:conf/aaai/DengHWLY19}, pair-wise loss  \cite{DBLP:conf/uai/RendleFGS09,DBLP:journals/kbs/LiuWZ18} and list-wise loss \cite{DBLP:conf/recsys/ShiLH10,DBLP:conf/recsys/ShiKBLOH12}. In this paper, we explore the point-wise loss only and leave the pair-wise and list-wise loss as a future work. Point-wise loss has been widely studied in collaborative filtering with explicit feedback under regression framework. The most commonly used point-wise loss is the squared loss, which minimizes the difference between the predicted value $\hat y_{ui}$ and its target value $y_{ui}$.

\begin{equation}
    L = \sum_{(u,i) \in \mathcal{Y}^+ \cup \mathcal{Y}^-} w_{ij} (y_{ui} - \hat{y}_{ui})^2  
\end{equation}
where $\mathcal{Y}^+ = \{(u, i) | y_{ui} = 1\}$ denotes the set of observed interactions, $\mathcal{Y}^- = \{(u, i) | y_{ui} = 0\}$ denotes the sampled unobserved interactions, i.e., negative instances and $w_{ij}$ denotes the weight of training instance $(u,i)$.
However, the squared loss is not suitable for implicit feedback because the implicit data is discrete and binary. The target value $y_{ui}$ is 1 if $u$ has interacted with $i$, otherwise 0.

Following \cite{DBLP:conf/www/HeLZNHC17}, we adopt the \textit{binary cross-entropy loss} as the objective function, which views top-N recommendation problem with implicit feedback as a binary classification problem.
\begin{equation}
    L = - \sum_{(u, i) \in \mathcal{Y}^+ \cup \mathcal{Y}^-} y_{ui} \log \hat{y}_{ui} + (1 - y_{ui}) \log (1 - \hat{y}_{ui})
\end{equation}

\section{Related Work\label{section:related_work}}

\subsection{Matrix Factorization based Collaborative Filtering}
Matrix Factorization (MF) typically represents each user/item as a low-dimensional embedding vector. Let $\mathbf{p}_u$ and $\mathbf{q}_i$ denote the latent vector for user $u$ and item $i$ in a shared embedding space, respectively. The relevance score $y_{ui}$ between user $u$ and item $i$ is estimated by the inner product of $\mathbf{p}_u$ and $\mathbf{q}_i$:
\begin{equation}
    \hat y_{ui} = <\mathbf{p}_u, \mathbf{q}_i> = \mathbf{p}_u^T \mathbf{q}_i
\end{equation}

Different from traditional MF methods, NSVD \cite{paterek:improving} represents users based on the items that they have rated. Note that each item $i$ is associated with two latent vector $\mathbf{q}_i$ and $\mathbf{y}_i$. Formally, the preference score of user $u$ to item $i$ is predicted as:
\begin{equation}
    \hat y_{ui} = b_u + b_i + \mathbf{q}_i^T \underset{\mathrm{user}\ u's\ \mathrm{representation}}{
        \underbrace{
            \left(\lvert R(u) \rvert^{-\frac{1}{2}} \sum_{j \in R(u)} \mathbf{y_j} \right)
        }
    }
\end{equation}
where $b_u$ and $b_i$ denote the bias terms of user $u$ and item $i$, respectively; and $R(u)$ is the set of items rated by user $u$. However, a main drawback of NSVD is that two different users who have rated the same set of items with different ratings have same representation. 

To address this problem, SVD++ \cite{DBLP:conf/kdd/Koren08} is proposed for recommendation with explicit ratings, which estimates the relevance score between user $u$ and item $i$ as follows:
\begin{equation}
    \hat y_{ui} = \mu + b_u + b_i + \mathbf{q}_i^T
    \underset{\mathrm{user}\ u's\ \mathrm{representation}}{
        \underbrace{
            \left(\mathbf{p}_u + \left \vert N(u) \right \vert^{-\frac{1}{2}} \sum_{j \in N(u)} \mathbf{y}_j \right)
        }
    }
\end{equation}
where $\mu$ is the average rating over all items, $\mathbf{p}_u$ is the latent vector for user $u$ and $N(u)$ denotes the set of items for which $u$ provided an implicit preference.
The user latent vector is complemented by the sum 
$\left \vert N(u) \right \vert^{-\frac{1}{2}} \sum_{j \in N(u)} \mathbf{y}_j$, which represents the perspective of implicit feedback. In other words, SVD++ leverages historical interactions to supplement the user latent factor rather than directly represent the user.

\subsection{Neighborhood based Collaborative Filtering}
For top-N recommendation, Kabbur et al. \cite{DBLP:conf/kdd/KabburNK13} proposed FISM (short  for \textit{Factored Item Similarity Model}), which learns the item-item similarity matrix as a product of two low-dimensional latent factor matrices.
Formally, the predictive model of FISM is
\begin{equation}
    \label{FISM}
    \hat y_{ui} = b_u + b_i + \mathbf{p}_i^T 
    \underset{\mathrm{user}\ u's\ \mathrm{representation}}{
        \underbrace{
            \left(
                \frac{1}{\lvert \mathcal{R}^+_u \rvert^\alpha} \sum_{j \in \mathcal{R}^+_u \setminus i} \mathbf{q}_j
            \right)
        }
    }
\end{equation}
where $\alpha$ is a hyper-parameter controlling the normalization effect, $\mathbf{p}_i$ and $\mathbf{q}_j$ denote the embedding vector for item $i$ and $j$, respectively.
In Equation (\ref{FISM}), the term in bracket can be viewed as the user $u$'s representation, which is aggregated from the embeddings of the historical items of $u$. 
% The main drawback of FISM is that all historical items of a user contribute equally when obtaining the user’s representation.

% To tackle these limitations, He et al. \cite{DBLP:journals/tkde/HeHSLJC18} proposed NAIS (short for \textit{Neural Attentive Item Similarity}) model, which distinguishes the importance of historical items by attention mechanism \cite{DBLP:journals/corr/BahdanauCB14}. Formally, the predictive model of NAIS is as follows:
% \begin{equation}
%     \hat y_{ui} = \mathbf{p}_i^T 
%     \underset{\mathrm{user}\ u's\ \mathrm{representation}}{
%         \underbrace{
%             \left(
%                 \sum_{j \in \mathcal{R}_u^+ \setminus \{i\}} a_{ij} \mathbf{q}_j 
%             \right)
%         }
%     }
% \end{equation}
% where $a_{ij}$ is a trainable parameter that denotes the attention weight of item $j$ in contributing to user $u$'s representation when predicting $u$'s preference on target item $i$.

\subsection{Deep Learning based Collaborative Filtering}

Despite the effectiveness of above approaches, they have an inherent limitation in their model design. Specifically, they use a simple and fixed inner product as interaction function, which is insufficient to capture the complex user-item interactions in the low-dimensional latent space. 

Neural collaborative filtering (NCF) \cite{DBLP:conf/www/HeLZNHC17} is thus proposed to learn the user–item interaction function via a multi-layer perceptron (MLP). 
% DeepMF \cite{DBLP:conf/ijcai/XueDZHC17} employs MLP to learn users' (items') representation with user's (item's) rating vector as input and then use \textit{cosine similarity} as interaction function to compute the relevance scores for each user-item pair.
DeepCF \cite{DBLP:conf/aaai/DengHWLY19} fuses representation learning-based CF methods and interaction function learning-based CF methods and applies multi-hot encoding on the ID feature of user $u$'s interacted items $\mathcal{R}_u^+$ to represent user $u$ (analogously for the items).
J-NCF \cite{DBLP:journals/tois/ChenCCR19} feeds user's (item's) rating vector into MLP to learn user's (item's) feature vector and then concatenate them to feed into another MLP to learn the interaction function.

All these methods mentioned above build the embedding function with either ID or historical interactions only. As reported in \cite{DBLP:conf/sigir/Wang0WFC19}, these methods cannot yield satisfactory embeddings and have to rely on the interaction fucntion to make up for the deficiency of suboptimal embeddings.
For this reason, DELF \cite{DBLP:conf/ijcai/ChengSZH18} is proposed to jointly adopt both ID and historical interactions to model user and item. However, they only employ dual embeddings to learn four kinds of interaction functions for each user-item pair and don not combine two types of embeddings to get better user's (item's) representation. In fact, it can be seen as the fusion of four MLP models which take the concatenation of different types of embeddings as the input.

\section{Proposed Methods\label{section:proposed_method}}

In this section, we first present the \textbf{D}ual-embedding based \textbf{N}eural \textbf{C}ollaborative \textbf{F}iltering (DNCF) framework. Before diving into the technical details, we first introduce some basic notations.

Throughout the paper, we used bold uppercase letter to denote a matrix (e.g., $\mathbf{R}$), bold lowercase letter to denote a vector (e.g., $\mathbf{p}$) and lowercase letter to denote a scalar (e.g., $y$).

\begin{figure*}
    \centering
    \includegraphics[scale=0.3]{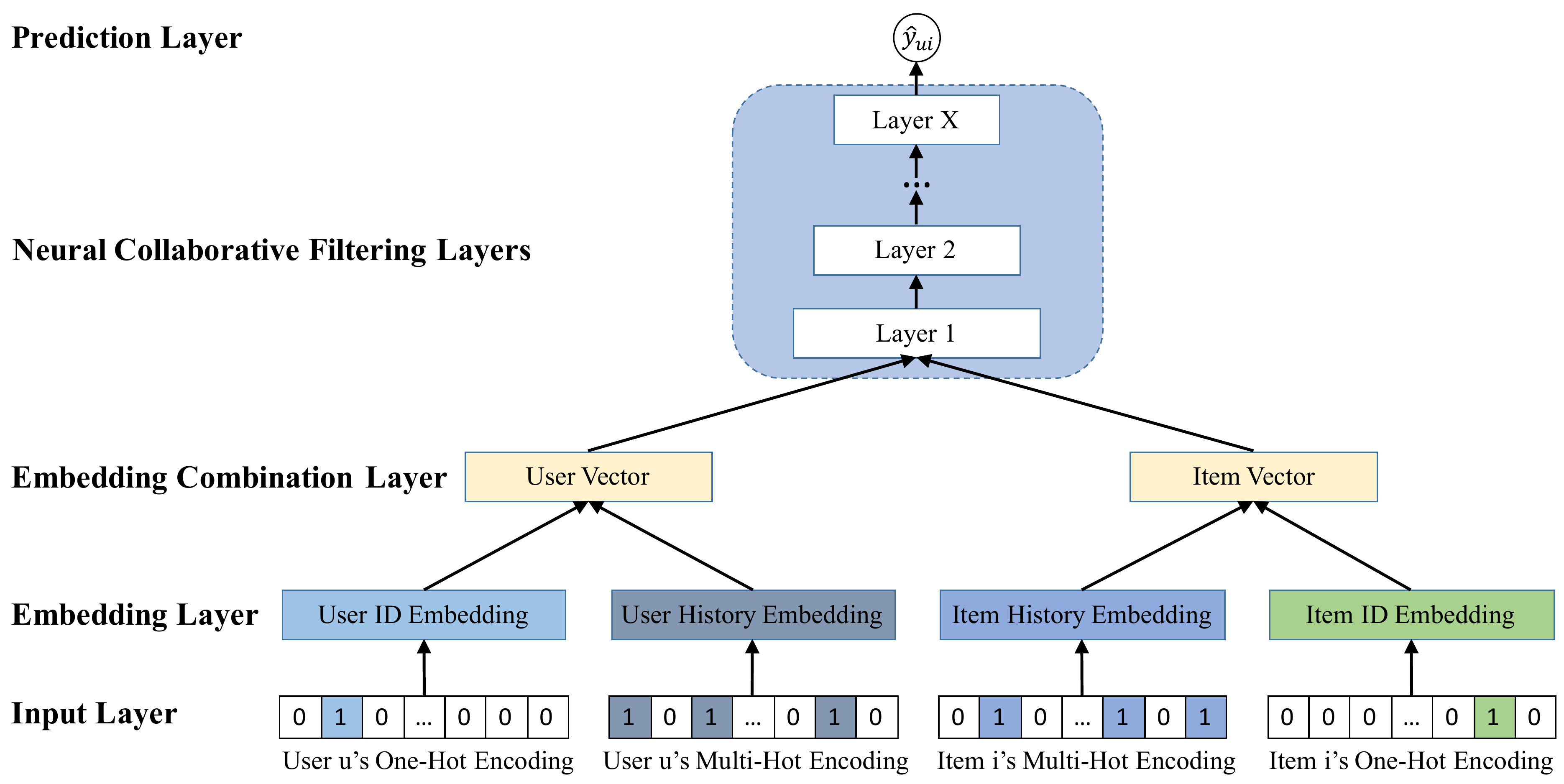}
    \caption{The architecture of DNCF.}
    \label{fig:architecture}
\end{figure*}

\subsection{DNCF Framework}
Figure \ref{fig:architecture} illustrates our proposed DNCF (short for Dual-embedding based Neural Collaborative Filtering) framework to model the interaction between users and items based on dual embeddings.

Next, we elaborate the architecture layer by layer. 

\textbf{Input Layer.}
We take ID and historical interactions as the input features and then transform them into binarized sparse vector with one-hot encoding and multi-hot encoding, respectively. Hence, we obtain two kinds of feature vectors for both user $u$ and $i$.

\textbf{Embedding Layer.}
The embedding layer projects each high-dimensional and sparse feature vector from the input layer into a low-dimensional and dense embedding vector. 
Let $\mathbf{P} \in \mathbb{R}^{M \times k}$, $\mathbf{Q} \in \mathbb{R}^{N \times k}$, $\mathbf{M} \in \mathbb{R}^{N \times k}$  and $\mathbf{N} \in \mathbb{R}^{M \times k}$ denote latent factor matrix for users and items from the perspective of ID and historical interactions, respectively, and $k$ is the dimension of embedding. The primitive embedding can be obtained as below:
\begin{equation}
    \mathbf{p}_u = \mathbf{P}^T \mathbf{x}_u, \quad
    \mathbf{q}_i = \mathbf{Q}^T \mathbf{x}_i
\end{equation}
where $\mathbf{p}_u$ and $\mathbf{q}_i$ denote the latent vector for user $u$ and item $i$ from the perspective of ID, respectively. We term it as \textit{ID embedding}.

As for the historical interactions, we learn the embedding by an aggregation function which summarizes the embeddings of the historical items (users) into a vector. We term it as \textit{history embedding}, short for \textit{historical interactions based embedding}. Take user $u$ as an example. Historical items are associated with another group of latent vectors $\mathbf{y}_j$
\begin{equation}
    \mathbf{m}_u = AGG(\{\mathbf{y}_j, \forall j \in \mathcal{R}_u^+\})
\end{equation}
where $\mathcal{R}_u^+$ is the set of items which user $u$ has interacted with, $\mathbf{y}_j$ is the corresponding column in $\mathbf{M}$ which represents items $j$, and $AGG(\cdot)$ denotes any aggregation function.

A common aggregation function is summation with normalization.
In this case, the embedding layer can be simplified as:
\begin{equation}
    \begin{aligned}
        \mathbf{m}_u &= \lvert \mathcal{R}_u^+ \rvert^{-\frac{1}{2}} \mathbf{M}^T \mathbf{Y}_{u*}, \quad
        \mathbf{n}_i = \lvert \mathcal{R}_i^+ \rvert^{-\frac{1}{2}} \mathbf{N}^T \mathbf{Y}_{*i}
    \end{aligned}
\end{equation}
where $\mathbf{m}_u$ and $\mathbf{n}_i$ denote the latent vector for user $u$ and item $i$ from the perspective of historical interactions, respectively; $\mathbf{Y}_{u*}$ and $\mathbf{Y}_{*i}$ are the $u$-th row and $i$-th column in $\mathbf{Y}$, respectively. 

\textbf{Embedding Combination Layer.}
The output of previous embedding layer is two kinds of embeddings for users and items, respectively. And the embedding combination layer integrates ID embedding with history embedding to get the final representation. Formally, it can be defined as:
\begin{equation}
    \mathbf{v}_u = g(\mathbf{p}_u, \mathbf{m}_u)
\end{equation}
where $g(\cdot)$ is any binary operator, which is termed as embedding combination function. Similarly, item $i's$ final representation $\mathbf{v}_i$ can be obtained.

\textbf{Neural Collaborative Filtering Layers.}
We feed the users' and items' representation into neural collaborative filtering layers to learn the interactions between users and items. Formally, this process is formulated as:
\begin{equation}
    \mathbf{z}_{out} = ncf(\mathbf{v}_u, \mathbf{v}_i)
\end{equation}
where $\mathbf{z}_{out}$ denotes the output vector of neural collaborative filtering layers $ncf(\cdot)$.

\textbf{Prediction Layer.}
The prediction layer maps the output vector of the neural collaborative filtering layers into the prediction score $\hat y_{ui}$ of the interaction between user $u$ and item $i$.

\subsection{Dual-embedding based Multi-Layer Perceptron}
Following the MLP model \cite{DBLP:conf/www/HeLZNHC17}, we propose Dual-embedding based Multi-Layer Perceptron (DMLP) which employs vector concatenation as embedding combination function $g(\cdot)$ and uses MLP to learn the interaction between user and item latent vector. Formally, the formulation of DMLP are given as follows:
\begin{equation}
    \begin{aligned}
        \mathbf{z}_0 &= (\mathbf{p}_u \oplus \mathbf{m}_u) \oplus (\mathbf{q}_i \oplus \mathbf{n}_i) \\
        \mathbf{z}_1 &= a_1 (\mathbf{W}^T_1 \mathbf{z}_0 + \mathbf{b}_1) \\
        &\cdots \\
        \mathbf{z}_{out} &= \mathbf{z}_L = a_L (\mathbf{W}^T_L \mathbf{z}_{L - 1} + \mathbf{b}_L) \\
        \hat y_{ui} &= \sigma(\mathbf{h}^T \mathbf{z}_{out} + b_{out})
    \end{aligned}
\end{equation}
where $\oplus$ denotes the concatenation operation between two vectors; $\mathbf{W}_l$, $\mathbf{b}_l$, $a_l$ and $\mathbf{z}_l$ denote the weight matrix, bias vector, activation function, and output vector of the $l$-th hidden layer; $\mathbf{h}$ and $b_{out}$ denote the weight vector and bias term of the prediction layer; $\sigma(\cdot)$ is the sigmoid function defined as $\sigma(x) = \frac{1}{1 + \exp(-x)}$. In this work, we choose \textit{Rectifier Linear Unit (ReLU)} as the activation function.

\subsection{Dual-embedding based Generalized Matrix Factorization}

Following the GMF model \cite{DBLP:conf/www/HeLZNHC17}, we propose 
Dual-embedding based Generalized Matrix Factorization (DGMF) which employs element-wise product to combine users' and items' representation.
\begin{equation}
        \phi(\mathbf{v}_u, \mathbf{v}_i) = g(\mathbf{p}_u, \mathbf{n}_u) \odot g(\mathbf{q}_i, \mathbf{n}_i)
        \label{equation:dgmf}
\end{equation}
where $\odot$ denotes the element-wise product of vectors.

Inspired by \cite{DBLP:conf/cikm/ZhangACC17}, we try stacking non-linear layer after element-wise product operation. However, it does not achieve better performance than the design of GMF \cite{DBLP:conf/www/HeLZNHC17} which has no hidden layer. This is probably because simple element-wise product is sufficient enough to capture the interactions between users and items in DGMF. For this reason, we directly project the vector into the predicted score:
\begin{equation}
    \hat y_{ui} = \sigma(\mathbf{h}^T \phi(\mathbf{v}_u, \mathbf{v}_i) + b_{out})
\end{equation}

To test the impact of $g(\cdot)$ in Equation (\ref{equation:dgmf}), we investigate four methods to combine the different embeddings into final user (item) vector for DGMF: \textit{element-wise sum}, \textit{element-wise mean}, \textit{concatenation} and \textit{attention mechanism} \cite{DBLP:journals/corr/BahdanauCB14}.
Take attention mechanism as an example:
\begin{equation}
    \begin{aligned}
        att(\mathbf{x}) &= \mathbf{h}^T_a ReLU(\mathbf{W}^T_a \mathbf{x} + \mathbf{b}_a), \\
        \alpha &= \frac{\exp(att(\mathbf{p}_u))}{\exp(att(\mathbf{p}_u)) + \exp(att(\mathbf{m}_u))} \\
        \mathbf{v}_u &= \alpha \mathbf{p}_u + (1 - \alpha) \mathbf{m}_u
    \end{aligned}
\end{equation}
where $\mathbf{W}_a \in \mathbb{R}^{k\times k^\prime}$ and $\mathbf{b}_a \in \mathbb{R}^{k^\prime}$ denote the weight matrix and bias vector of the attention network, respectively, and $k^\prime$ denotes the size of hidden layer; $\mathbf{h}_a \in \mathbb{R}^{k^\prime}$ denotes the weight vector of the output layer of the attention network. Likewise, we can also get the final item representation.

Without special mention, we use simple element-wise sum as embedding combination function $g(\cdot)$. And we compare the impact of these methods for DGMF and the experimental results are shown in section \ref{section:dgmf_variants}.

\subsection{Fusion of DGMF and DMLP}
Following the design of NeuMF \cite{DBLP:conf/www/HeLZNHC17}, we allow DGMF and DMLP to learn separate embeddings and combine the two models by concatenating the output vectors of neural CF layers. And then we feed them into a fully connected layer. Specifically, it can be formulated as:
\begin{equation}
    \begin{aligned}
        \phi^{DGMF} &= g(\mathbf{p}_u^G,\mathbf{m}_u^G) \odot g(\mathbf{q}_i^G,\mathbf{n}_i^G) \\
        \phi^{DMLP} &= MLP(\mathbf{p}_u^M \oplus \mathbf{m}_u^M \oplus \mathbf{q}_i^M \oplus \mathbf{n}_i^M) \\
        \hat y_{ui} &= \sigma(\mathbf{h}^T(\phi^{DGMF} \oplus \phi^{DMLP}) + b_{out})
    \end{aligned}
\end{equation}
where $\mathbf{p}_u^G$ and $\mathbf{p}_u^M$ denote the user's ID embedding for DGMF and DMLP, respectively, and similar notations for others. We refer to this model as DNMF, short for \textit{Dual-embedding based Neural Matrix Factorization}.

\subsubsection{Pre-training}

As reported in \cite{DBLP:conf/ijcai/0001DWTTC18}, the initialization plays a significant role for the convergence and performance of deep learning model. Since DNMF is an ensemble of DGMF and DMLP, we propose to initialize DNMF using the pre-trained models of DGMF and DMLP. 
First, we train DGMF and DMLP from scratch using Adam \cite{DBLP:journals/corr/KingmaB14} until convergence.
Then, we use their model parameters as the initialization for the corresponding parts of DNMF's parameters.
Notice that the DNMF with pre-training is optimized by the vanilla SGD rather than Adam. This is because Adam requires momentum information to update parameters which is not saved in DNMF with pre-training.

\subsection{Model Analysis}
In this subsection, we first show how DNCF generalizes SVD++ \cite{DBLP:conf/kdd/Koren08} and FISM \cite{DBLP:conf/kdd/KabburNK13}. In what follows, we analyze the time complexity of DNMF.

\subsubsection{DNCF Generalizes SVD++ $\&$ FISM}
Both SVD++ and FISM can be viewed as a special case of DNCF. In particular, we use element-wise sum as embedding combination function. In the neural collaborative filtering layers, we employ inner product to model the interactions between users and items.
We term this model as \textit{DNCF-MF}, which can be formulated as:
\begin{equation}
    \hat y_{ui} = 
    (\mathbf{p}_u + 
    \left \vert \mathcal{R}^+_u \right \vert^{-\frac{1}{2}}
    \sum_{i \in \mathcal{R}^+_u} \mathbf{y}_i)^T 
    (\mathbf{q}_i + 
    \left \vert \mathcal{R}^+_i \right \vert^{-\frac{1}{2}}
    \sum_{u \in \mathcal{R}^+_i} \mathbf{y}_u)
    \label{equation:DNCF-mf}
\end{equation}

Clearly, by disabling additional embeddings for items which aggregates from the embedding of the historical users, we can exactly recover SVD++ model. Analogously, if we disable primitive user embeddings and additional embedding for items in Equation (\ref{equation:DNCF-mf}), we can recover FISM.

\subsubsection{Time Complexity Analysis}
For the embedding layer, the matrix multiplication has computational complexity $O\left((k + d_0)(\lvert \mathcal{R}^+_u \rvert + \lvert \mathcal{R}^+_i \rvert)\right)$, where $k$ denotes the embedding size for DGMF part which is equal to the number of predictive factors, $d_0$ represents the embedding size for DMLP part and $\lvert \mathcal{R}^+_u \rvert$ denotes the number of historical items interacted by user $u$ and similar notations for $\lvert \mathcal{R}^+_i \rvert$. For the collaborative filtering layer, the time complexity is $O(\sum_{l=1}^L d_l d_{l-1})$, where $d_l$ represents the size of the $l$-th hidden layer and $d_L = k$. The prediction layer only involves inner product of two vectors which can be done in $O(d_L)$. Therefore, the overall time complexity for evaluating a prediction with DNMF is 
$
O\left(
    (k + d_0)(\lvert \mathcal{R}^+_u \rvert + \lvert \mathcal{R}^+_i \rvert)
    + \sum_{l=1}^L d_l d_{l-1}
    \right )
$.

\section{Experiments\label{section:experiment}}
In this section, we conduct plenty of experiments on four publicly accessible datasets to answer the following research questions:

\textbf{RQ1} Do our proposed DNCF methods outperform the state-of-the-art collaborative filtering methods?

\textbf{RQ2} How do the key hyper-parameter settings impose influence on the performance of our DNCF approaches?

\textbf{RQ3} Are deeper layers of hidden units helpful for the recommendation performance of DNCF?

\textbf{RQ4} How is the performance of DNCF impacted by different embedding combine functions?

Hereinafter, we first describe experimental settings and then answer the above questions one by one.

\subsection{Experimental Settings}
\textbf{Dataset Description.} 
We evaluate our model in four real-world datasets: MovieLens 1M, Last.FM, AMusic and AToy. The statistics of the four datasets are summarized in Table \ref{table:statistics-dataset}. Following previous work \cite{DBLP:conf/aaai/DengHWLY19}, we use the processed datasets\footnote{The processed datasets are downloaded from: https://github.com/familyld/DeepCF}.

\begin{table}[htbp]
	\caption{Statistics of the Datasets}
	\centering
	\resizebox{0.5\textwidth}{!}
	{
	\begin{tabular}{|c|c|c|c|c|}
		\hline
		\textbf{Dataset} & \textbf{\#Users} & \textbf{\#Items} & \textbf{\#Interactions} & \textbf{Density} \\
		\hline
		MovieLens 1M & 6,040 & 3,706 & 1,000,209  & 4.47\% \\
		\hline
		Last.FM & 1,741 & 2,665 & 69,149 & 1.49\% \\
		\hline
		AMusic & 1,776 & 12,929 & 46,087 & 0.20\% \\
		\hline
		AToy & 3,137 & 33,953 & 84,642 & 0.08\% \\
		\hline
	\end{tabular}
	}
	\label{table:statistics-dataset}
\end{table}

\textbf{Evaluation Protocols.}
 Following \cite{DBLP:conf/www/HeLZNHC17, DBLP:conf/ijcai/XueDZHC17, DBLP:conf/cikm/BaiWZZ17, DBLP:conf/ijcai/ChengSZH18, DBLP:conf/aaai/DengHWLY19}, we adopted \textit{leave-one-out} evaluation which holds out the latest interaction of each user as the test set and uses the remaining interactions for training. In terms of evaluation metrics, we used \textit{Hit Ratio} at rank k (HR@k) \cite{DBLP:conf/kdd/KabburNK13} and \textit{Normalized Discounted Cumulative Gain} at rank k (NDCG@k) \cite{DBLP:conf/www/HeLZNHC17, DBLP:conf/cikm/BaiWZZ17, DBLP:journals/tois/XueHWXLH19, DBLP:conf/ijcai/ChengSZH18, DBLP:conf/aaai/DengHWLY19} to evaluate the performance of the ranked list generated by our models. 
In this case, HR@k is defined as
\begin{equation}
	HR@k = 
	\begin{cases}
	1, & \mathrm{if\ the\ test\ item\ is\ in\ the\ top\ k} \\
	0, & \mathrm{otherwise}
	\end{cases}
\end{equation}
And NDCG@k is defined as
\begin{equation}
	NDCG@k = \frac{1}{\log_2 (pos_i + 1)}
\end{equation}
where $pos_i$ denotes the position of the test item in the ranked recommendation list for the $i$-th hit.

Unless otherwise stated, the ranked list is truncated at 10 for both metrics. The metric of HR@10 is capable of measuring intuitively if the test item is present at the top-10 ranked list and NDCG@10 illustrates the quality of ranking which assigns higher score to hits at top position ranks \cite{DBLP:conf/www/HeLZNHC17}. We calculated both metrics for each test user and reported the average score. 

\textbf{Baselines.}
To evaluate the performance of our proposed model, we compared it with the following approaches:

\begin{itemize}
    \item \textbf{ItemPop}.
    This is a non-personalized method that is often used as a benchmark for recommendation tasks. Items are ranked by their popularity measured by the number of interactions.
    \item \textbf{eALS} \cite{DBLP:conf/sigir/HeZKC16}. This is a state-of-the-art MF method which learns MF model by optimizing a point-wise regression loss that treats all missing data as negative feedback with a smaller weight\footnote{\url{https://github.com/hexiangnan/sigir16-eals}}.
    \item \textbf{BiasedMF} \cite{DBLP:conf/recsys/RendleKZA20}. This method optimizes biased MF model with binary cross-entropy loss to learn from implicit feedback data\footnote{\url{https://github.com/google-research/google-research/tree/master/dot_vs_learned_similarity}}.
    \item \textbf{GMF} \cite{DBLP:conf/www/HeLZNHC17}. This is a generalized version of MF which extends MF by introducing non-linear activation function and allowing varying importance of latent dimensions\footnote{\url{https://github.com/hexiangnan/neural_collaborative_filtering}\label{note:ncf}}.
    \item \textbf{MLP} \cite{DBLP:conf/www/HeLZNHC17}.
    This approach applies the one-hot encoding of users' (items') ID to represent users (items) and adopts multi-layer perceptron instead of the fixed inner product to learn the non-linear interactions between users and items\textsuperscript{\ref{note:ncf}}.
    \item \textbf{NeuMF} \cite{DBLP:conf/www/HeLZNHC17}.
    This is a state-of-the-art interaction function learning-based MF model which combines the last hidden layer of GMF and MLP to learn the interaction function based on binary cross-entropy loss\textsuperscript{\ref{note:ncf}}.
    \item \textbf{CFNet-ml} \cite{DBLP:conf/aaai/DengHWLY19}.
    This is a interaction function learning-based CF method which employs historical interactions as the input of the model and then feeds them into MLP to learn the complex interactions between users and items \footnote{\url{https://github.com/familyld/DeepCF}\label{note:deepcf}}.
    \item \textbf{CFNet} \cite{DBLP:conf/aaai/DengHWLY19}.
    This is a state-of-the-art method which combines the strengths of representation learning-based and matching function learning-based CF method\textsuperscript{\ref{note:deepcf}}.
    \item \textbf{J-NCF} \cite{DBLP:journals/tois/ChenCCR19}.
    This is a state-of-the-art method which applies a joint neural network that couples deep feature learning and deep interaction modeling with a rating matrix. For a fair comparison, we choose \textit{binary cross-entropy loss function} as objective function.
    We employ three layers in the DF network with the size of [256,128,64] and two layers in the DI network with the size of [128,64].
\end{itemize}

As our proposed methods focus on modeling the relationship between users and items, we mainly compare with user–item models. We do not compare with DELF \cite{DBLP:conf/ijcai/ChengSZH18} because its performance is similar to or worse than NeuMF.

\textbf{Parameter Settings.}
We implemented our proposed model based on Keras\footnote{\url{https://keras.io}} and Tensorflow\footnote{\url{https://www.tensorflow.org}}, which will be released publicly upon acceptance. To determine hyper-parameters of DNCF methods, we held-out the latest interaction for each user in the training set as the validation data and tuned hyper-parameters on it. We sampled 4 negative instances per positive instance. For DGMF and DMLP, we randomly initialized model parameters with a Gaussian Distribution (with a mean of 0 and standard deviation of 0.01), optimizing the model with mini-batch Adam \cite{DBLP:journals/corr/KingmaB14}.
We used the batch size of 256 and the learning rate of 0.001. And the regularization coefficient $\lambda$ is set to $1e^{-6}$. The size of last hidden layer was referred to as \textit{predictive factors} \cite{DBLP:conf/www/HeLZNHC17} and we evaluated the factors of [8,16,32,64]. Unless specified, we employed three hidden layers for MLP. For instance, if the size of predictive factors is 64, neural collaborative filtering layers follow $256 \rightarrow 128 \rightarrow 64$ and the embedding size is 64.

\subsection{Performance Comparison (RQ1)}

\begin{table*}[tb]
    \caption{Performance of HR@10 and NDCG@10 of different methods at predictive factor 64.}
    \label{tab:performance_comparison}
    \centering
    \resizebox{\textwidth}{!}
    {
    \begin{tabular}{|c|cc|cc|cc|cc|}
        \hline
        \textbf{Datasets} & 
        \multicolumn{2}{|c|}{\textbf{MovivLens 1M}} & \multicolumn{2}{|c|}{\textbf{Last.FM}} & 
        \multicolumn{2}{|c|}{\textbf{AMusic}} & 
        \multicolumn{2}{|c|}{\textbf{AToy}}\\
        \hline
        \textbf{Methods} & \textbf{HR@10} & \textbf{NDCG@10}
        & \textbf{HR@10} & \textbf{NDCG@10} 
        & \textbf{HR@10} & \textbf{NDCG@10}
        & \textbf{HR@10} & \textbf{NDCG@10} \\
        \hline \hline
        \textbf{ItemPop} & 0.4535 & 0.2542 & 0.6628 & 0.3862 & 0.2483 & 0.1304 & 0.2840 & 0.1518\\
        \hline
        \textbf{eALS} & 0.7018 & 0.4280 & 0.8265 & 0.5162 & 0.3711 & 0.2352 & 0.3717 & 0.2434\\
        \hline
        \textbf{BiasedMF} & \textbf{0.7295} & \textbf{0.4492} & \textbf{0.9041} & 0.6170 & 0.3846 & 0.2384 & 0.3711 & 0.2297\\ 
        \hline \hline
        \textbf{GMF} & 0.7026 & 0.4248 & 0.8759 & 0.5981 & 0.3502 & 0.2135 & 0.3800 & 0.2291\\
        \hline
        \textbf{MLP} & 0.6950 & 0.4171 & 0.8604 & 0.5658 & 0.3941 & 0.2274 & 0.3825 & 0.2283\\ 
        \hline
        \textbf{NeuMF} & 0.7172 & 0.4380 & 0.8874 & 0.6068 & 0.3992 & 0.2370 & 0.4017 & 0.2505\\
        \hline \hline
        \textbf{CFNet-ml} & 0.7075 & 0.4265 & 0.8811 & 0.5860 & 0.4071 & 0.2420 & 0.3931 & 0.2293\\
        \hline
        \textbf{CFNet} & 0.7253 & 0.4416 & \textbf{0.9064} & \textbf{0.6270} & 0.4116 & \textbf{0.2601} & \textbf{0.4150} & \textbf{0.2513}\\
        \hline
        \textbf{J-NCF} & 0.7023 & 0.4232 & 0.8748 & 0.5909 & 0.4099 & 0.2360 & 0.3672 & 0.2041\\ 
        \hline \hline
        \textbf{DGMF} & 0.7232 & 0.4440 & 0.8914 & 0.6197 & 0.4200 & 0.2584 & 0.3994 & 0.2500\\
        \hline 
        \textbf{DMLP} & 0.7215 & 0.4468 & 0.8817 & 0.5948 & \textbf{0.4257} & 0.2558 & 0.3985 & 0.2353\\
        \hline
        \textbf{DNMF} & \textbf{0.7341} & \textbf{0.4531} & 0.9006 & \textbf{0.6257} & \textbf{0.4358} & \textbf{0.2712} & \textbf{0.4182} & \textbf{0.2645}\\
        \hline
    \end{tabular}
    }
\end{table*}

\begin{figure*}[htbp]
    \centering
    \subfigure[MovieLens 1M --- HR@10]{
        \includegraphics[scale=0.25]{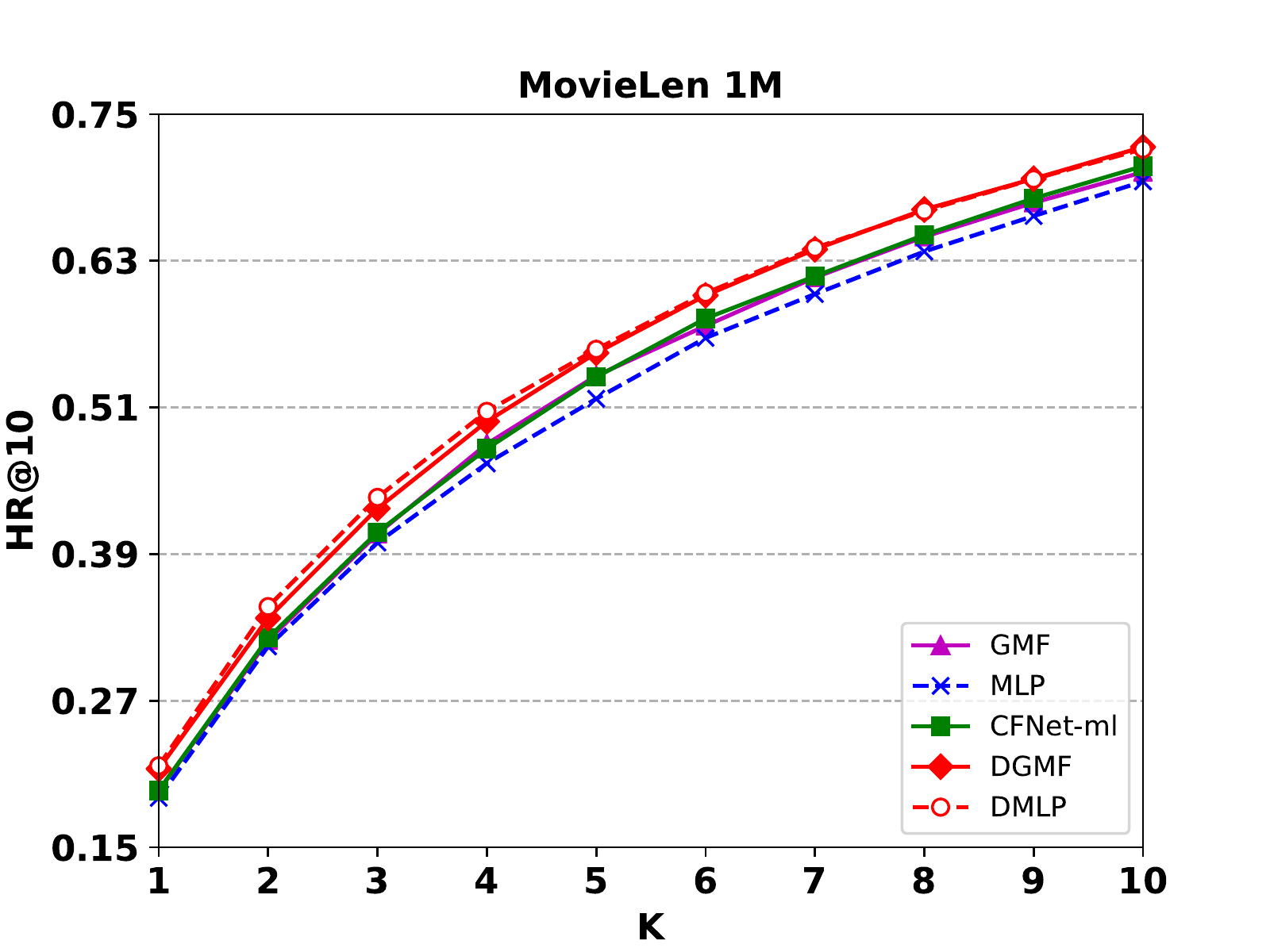}
    }
    \subfigure[MovieLens 1M --- NDCG@10]{
        \includegraphics[scale=0.25]{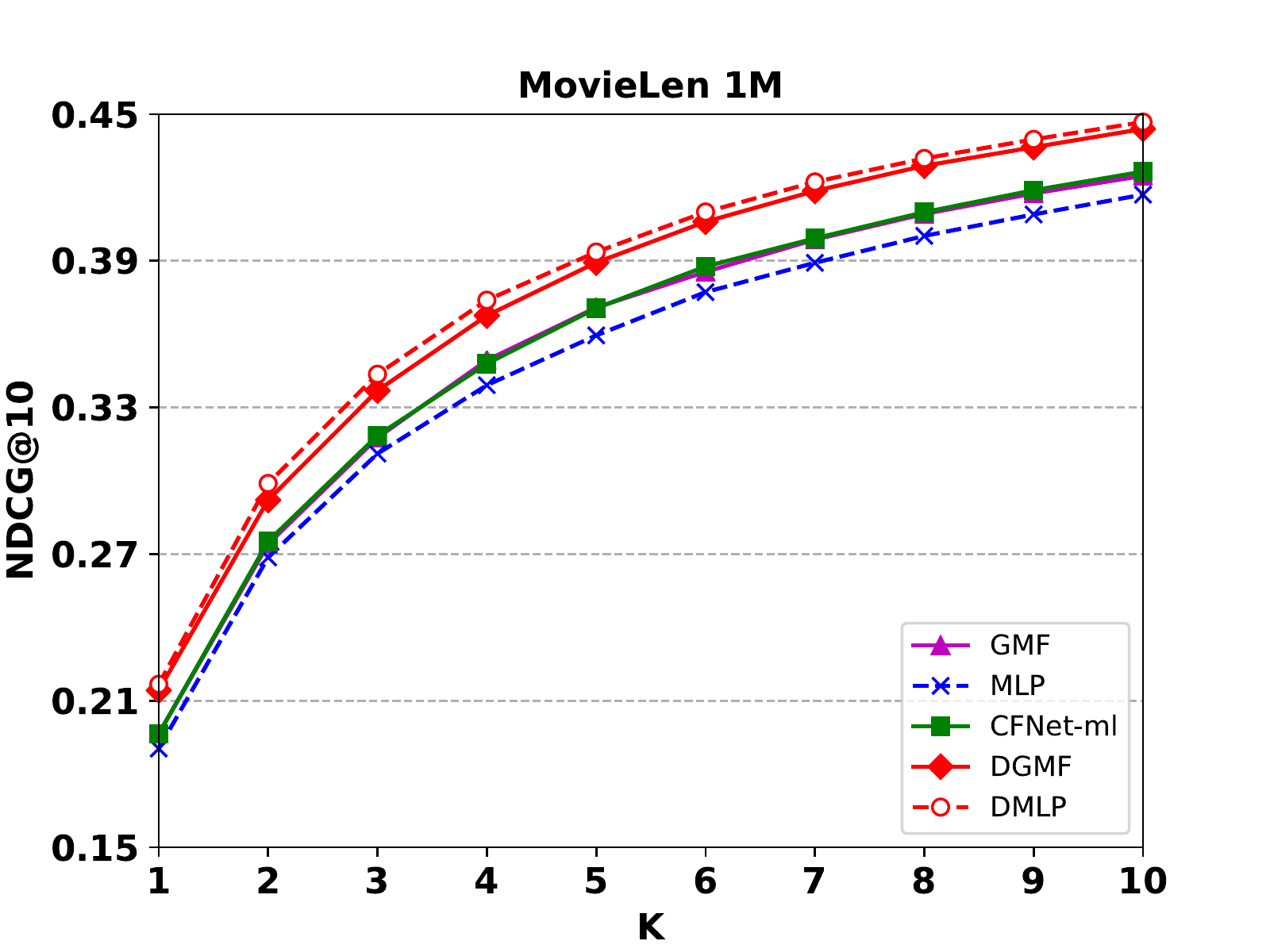}
    }
    \subfigure[Last.FM --- HR@10]{
        \includegraphics[scale=0.25]{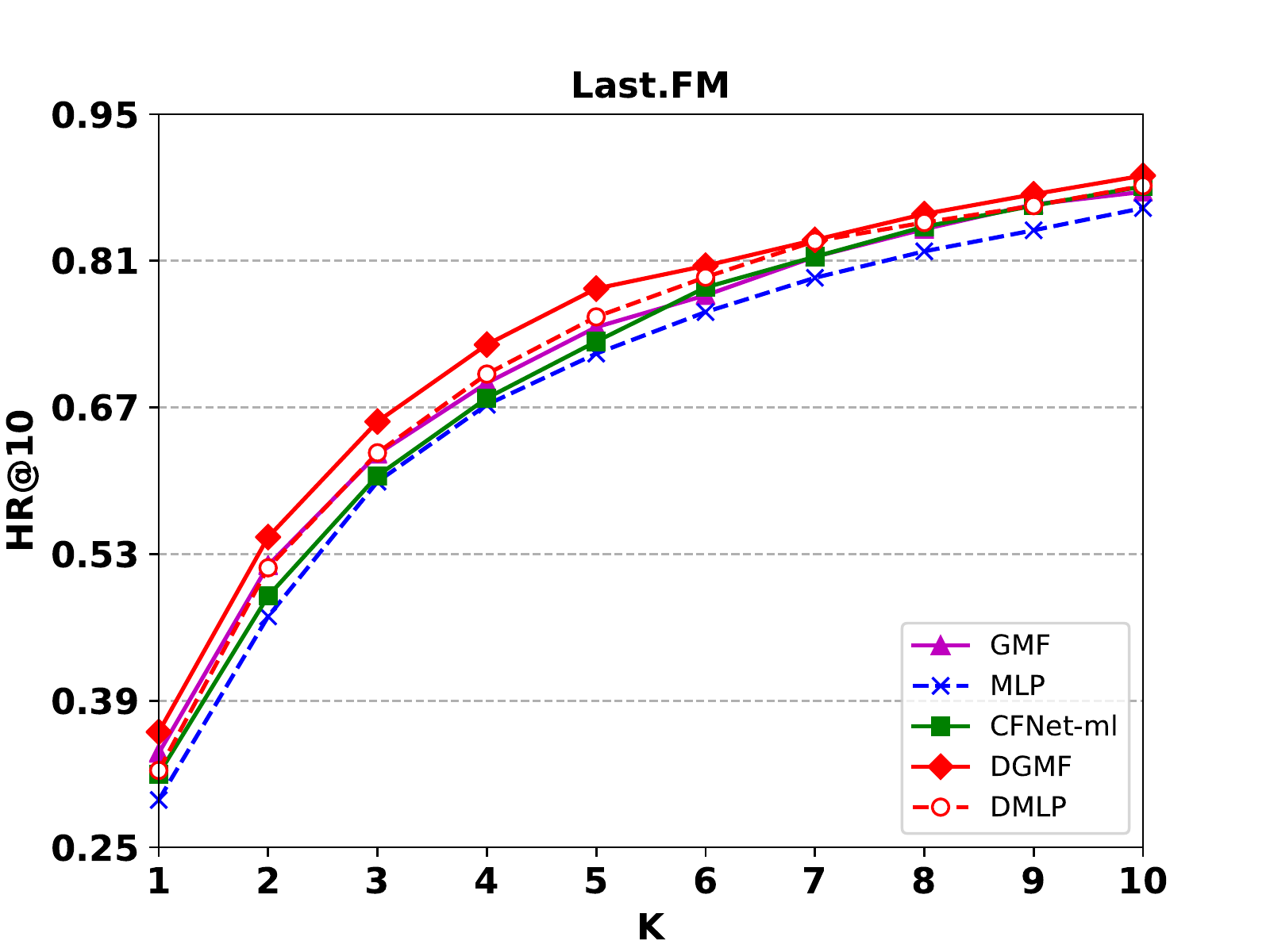}
    }
    \subfigure[Last.FM --- NDCG@10]{
        \includegraphics[scale=0.25]{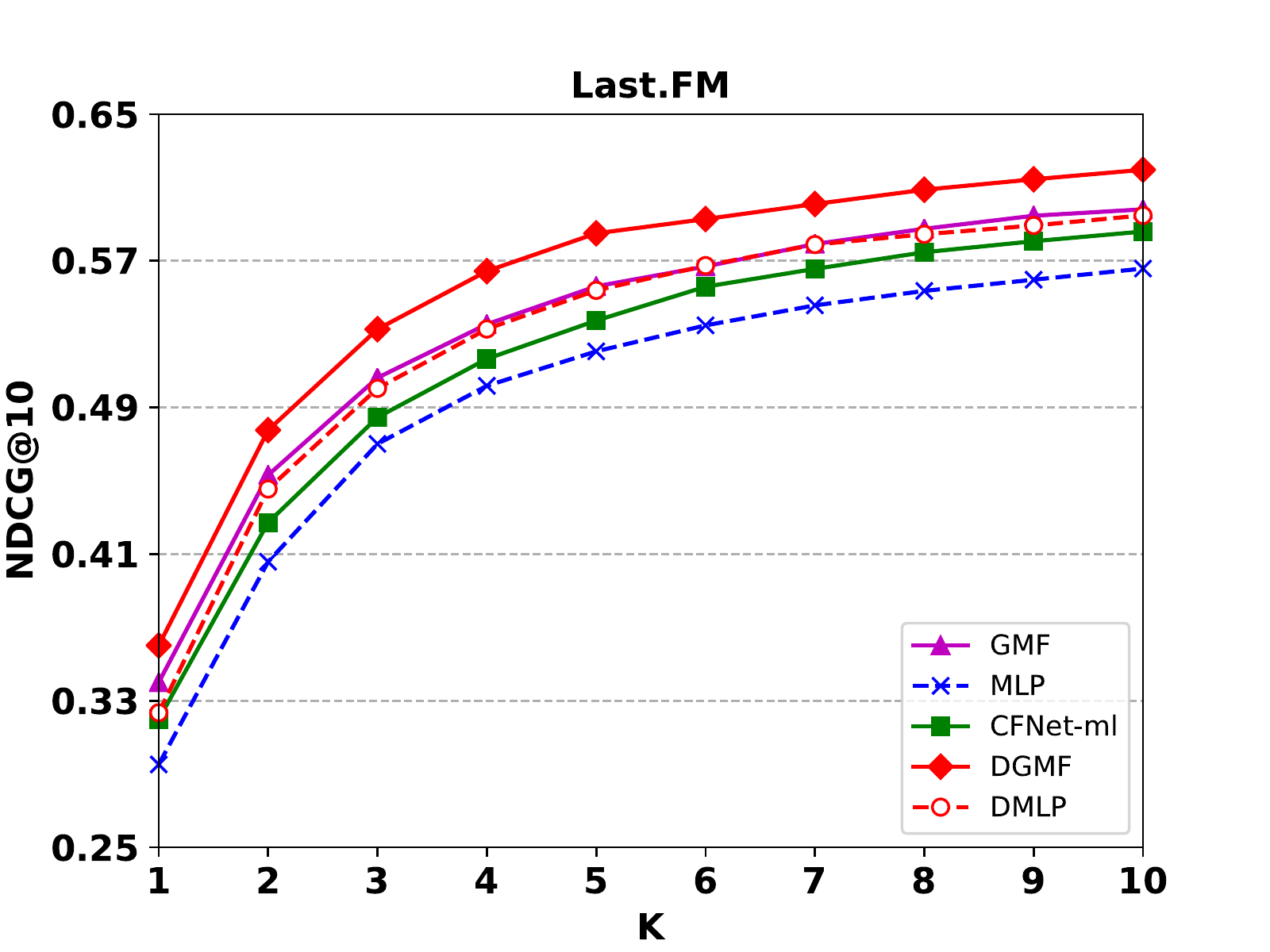}
    }
    \caption{Evaluation of Top-K item recommendation where K ranges from 1 to 10 on the MovieLens 1M and Last.FM.}
    \label{fig:topk}
\end{figure*}

Table \ref{tab:performance_comparison} shows the performance of HR@10 and NDCG@10 of all compared methods. The best and the second best results are highlighted as bold font. For a fair comparison, the size of predictive factors is fixed to 64 for all methods. For eALS and BiasedMF, the number of predictive factors is equal to the number of latent factors.

We have the following observations:
\begin{itemize}
    \item DNMF yields the best performance on most of the datasets except the Last.FM dataset. Specifically, DNMF improves over the strongest baselines by 0.7\%, 5.1\% and 3.0\% for MovieLens 1M, AMUsic and AToy, respectively. On the Last.FM dataset, DNMF slightly underperforms BiasedMF and CFNet in terms of HR@10 while outperforms other baseline methods except CFNet in terms of NDCG@10. This result justifies the effectiveness of our proposed DNCF framework that models the interactions between users and items based on dual embeddings.

    \item DMLP outperforms MLP and CFNet-ml by a large margin. Besides, DGMF also demonstrates consistent improvements over GMF. This findings provide empirical evidence for the effectiveness of utilizing historical interactions based embedding to augment the representation. For baseline methods, CFNet-ml achieves better performance than MLP on all datasets. This indicates adopting historical interactions as input can get better representation than ID. Between DMLP and DGMF, DMLP slightly underperforms DGMF, which is in consistent with the results shown in \cite{DBLP:conf/www/HeLZNHC17}.

    \item J-NCF underperforms CFNet on all datasets. One reason is that the superiority of J-NCF might mainly be attributed to its loss function. In original paper \cite{DBLP:journals/tois/ChenCCR19}, the authors proposed a hybrid loss function which combines point-wise and pair-wise loss function.
\end{itemize}

We also evaluate the performance of Top-K recommended lists where the ranking position K ranges from 1 to 10 on the MovieLens 1M and Last.FM as illustrated in Figure \ref{fig:topk}. To make the figure more clear, we only show MLP, GMF and their variants --- CFNet-ml, DMLP and DGMF. We can find that DGMF achieves consistent improvements over GMF across positions on both datasets. Likewise, DMLP outperforms CFNet-ml and MLP on all ranking positions. This demonstrates the advantage of dual embeddings again. For baseline methods, MLP underperforms GMF and CFNet-ml on both datasets. For MovieLens 1M, CFNet-ml achieves better performance than GMF on both metrics. However, it underperfoms GMF on Last.FM in terms of NDCG.

\subsubsection{Utility of Pre-training}

To demonstrate the impact of pre-training for DNMF, we compared the performance of two versions of DNMF --- with and without pre-training. Different from the DNMF with pre-training, we used mini-batch Adam to learn the DNMF without pre-training with random initializations.
The experimental results are provided in Table \ref{tab:pretraining}. We can find that the DNMF with pre-training outperforms the DNMF without pre-training on all datasets. 
The relative improvements of DNMF with pre-training are 2.6\%, 1.8\%, 9.7\% and 5.9\% for MovieLens 1M, Last.FM, AMusic and AToy, respectively.
This result verifies the utility of pre-training process for DNMF.

\begin{table}[htb]
    \caption{Performance of DNMF with/without pre-training at predictive factor 64.}
    \centering
    \resizebox{0.5\textwidth}{!}{
        \begin{tabular}{|c|c|c|c|c|}
            \hline
            \multirow{2}{*}{\textbf{Datasets}} & 
            \multicolumn{2}{c|}{\textbf{Without pre-training}} & 
            \multicolumn{2}{c|}{\textbf{With pre-training}}\\
            \cline{2-5}
            & \textbf{HR@10} & \textbf{NDCG@10} & \textbf{HR@10} & \textbf{NDCG@10} \\
            \hline 
            \textbf{MovieLens 1M} & 0.7159 & 0.4412 & \textbf{0.7341} & \textbf{0.4531}\\
            \hline
            \textbf{Last.FM} & 0.8874 & 0.6122 & \textbf{0.9006} & \textbf{0.6257}\\
            \hline
            \textbf{AMusic} & 0.4015 & 0.2447 & \textbf{0.4358} & \textbf{0.2712}\\
            \hline
            \textbf{AToy} & 0.3950 & 0.2496 & \textbf{0.4182} & \textbf{0.2645}\\
            \hline
        \end{tabular}
    }
    \label{tab:pretraining}
\end{table}

\subsection{Sensitivity to Hyper-parameter (RQ2)}
In this section, we study the impact of different hyper-parameter values on the performance of our proposed models.

% Figure \ref{fig:alpha} shows the performance of HR@10 and NDCG@10 on MovieLens 1M and Last.FM with respect to the trade-off parameter $\alpha$. Results on AMusic and AToy show the same trend and thus they are omitted due to space limitation. We can see that the performances of HR and NDCG increase gradually with increase of $\alpha$ with step size of 0.1. For MovieLens 1M, the best HR@10 is obtained when the value of $\alpha$ is set to 0.7 while the best NDCG@10 is obtained when the value of $\alpha$ is equal to 0.5. While on Last.FM, the best result in terms of HR is obtained when the value of $\alpha$ is equal to 0.6 and for NDCG the value is 0.8. And then the performances are degraded drastically with the continuing increase of $\alpha$. To sum up, the optimal value of $\alpha$ is around 0.5 to 0.8. 

\subsubsection{Number of Predictive Factors} 

\begin{figure*}[htb]
    \centering
    \subfigure[MovieLens 1M --- HR@10]{
        \includegraphics[scale=0.25]{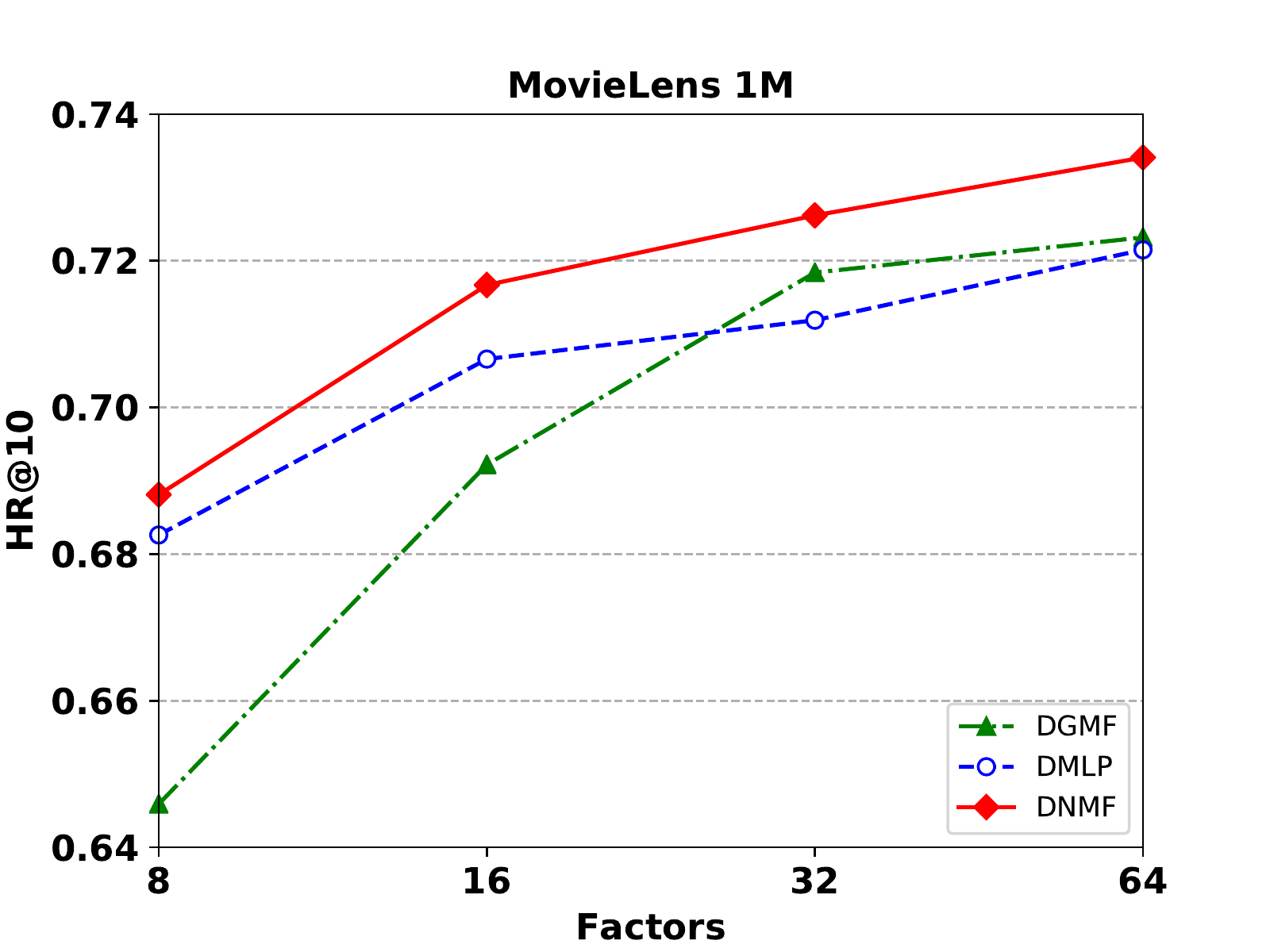}
    }
    \subfigure[MovieLens 1M --- NDCG@10]{
        \includegraphics[scale=0.25]{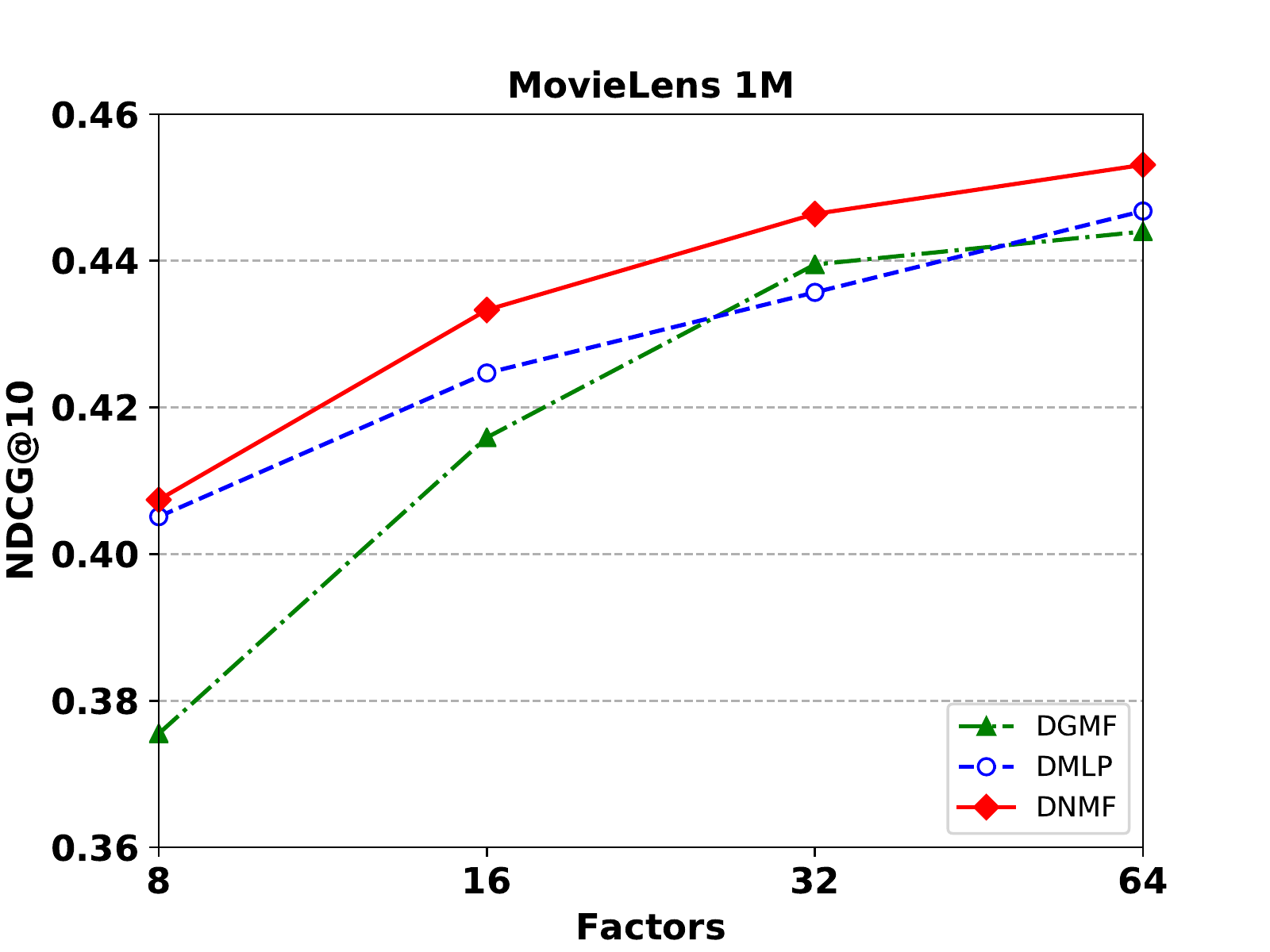}
    }
    \subfigure[Last.FM --- HR@10]{
        \includegraphics[scale=0.25]{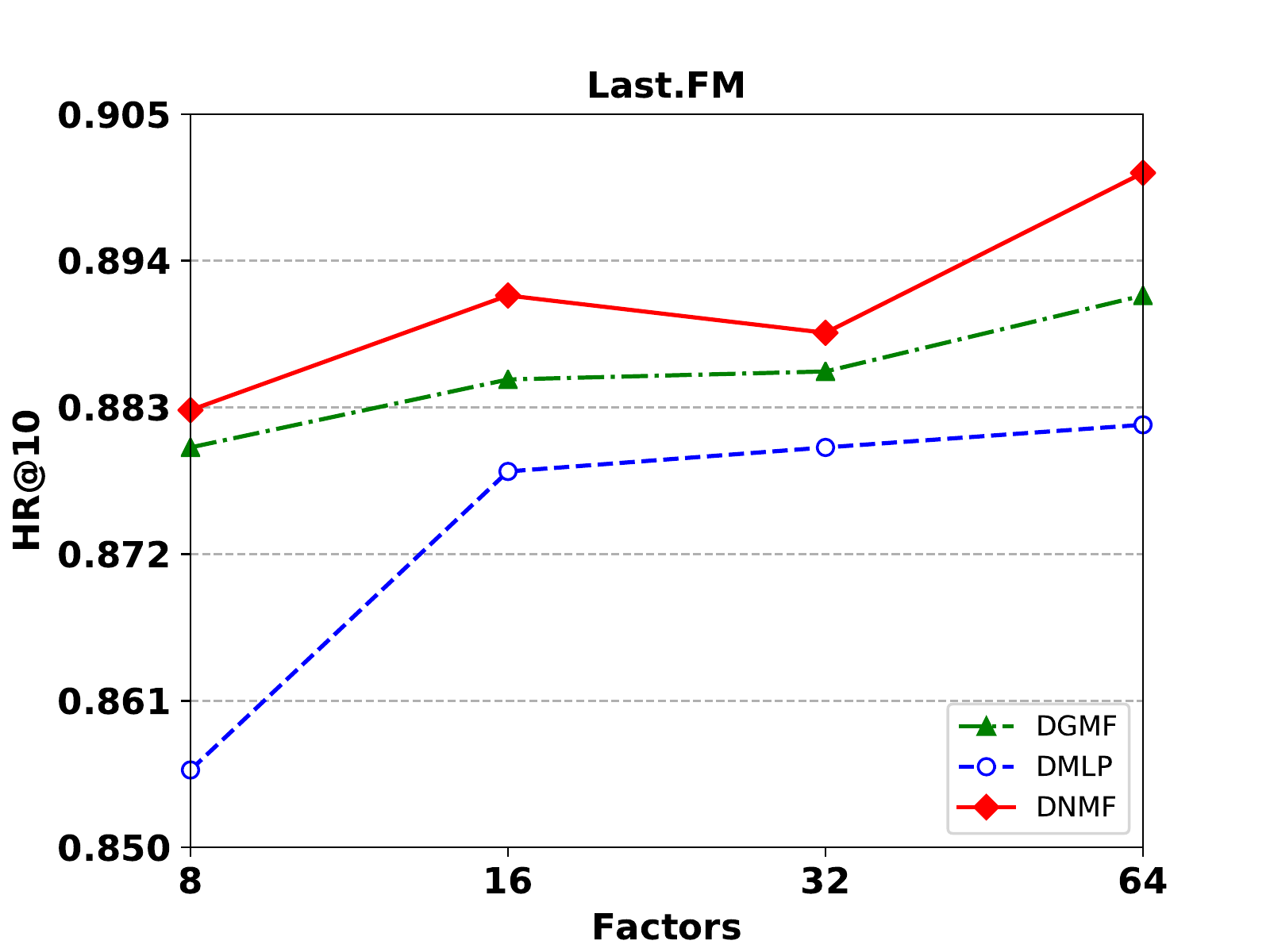}
    }
    \subfigure[Last.FM --- NDCG@10]{
        \includegraphics[scale=0.25]{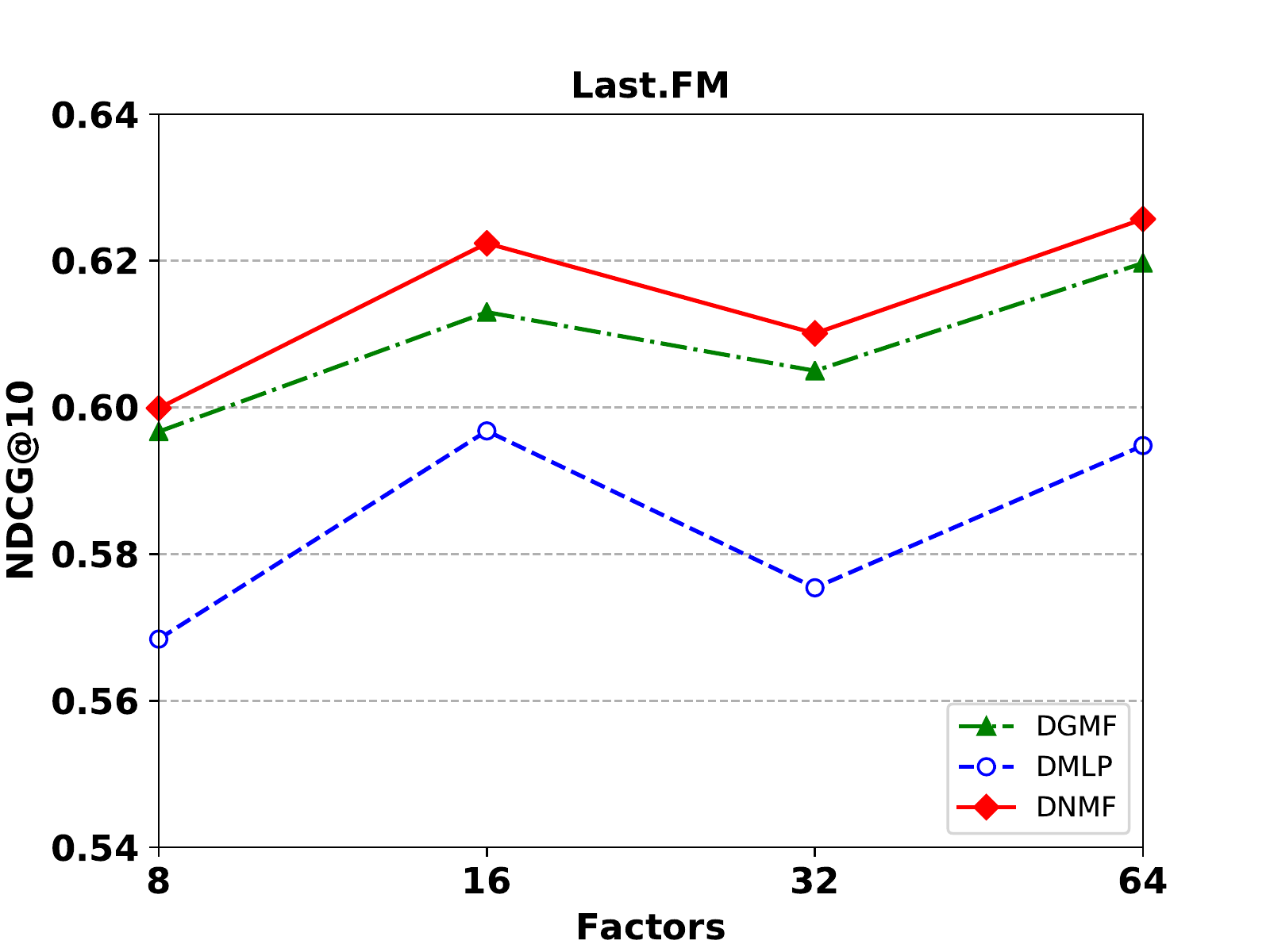}
    }
    \caption{Performance of HR@10 and NDCG@10 w.r.t. the number of predictive factors.}
    \label{fig:factors}
\end{figure*}
Fixing the remaining parameter values, we did a full parameter study for predictive factors. Figure \ref{fig:factors} shows the performance of HR@10 and NDCG@10 on MovieLens 1M and Last.FM with respect to the number of predictive factors. The proposed models offer the best performance with 64 predictive factors on both datasets. For MovieLens 1M, the performance of all models increase gradually with the increase of predictive factors. For Last.FM, the performances of NDCG@10 of DGMF and DMLP increase first and then decrease. 
It's worth noticing that for MovieLens 1M with a small predictive factors of 8 and 16, DGMF underperforms DMLP, while shows consistent improvement over DMLP on the Last.FM dataset. One possible reason is that DGMF has the ability to express stronger representation than DMLP on the relative sparse dataset.

\subsubsection{Negative Sampling Ratio}

\begin{figure*}[htb]
    \centering
    \subfigure[MovieLens 1M --- HR@10]{
        \includegraphics[scale=0.25]{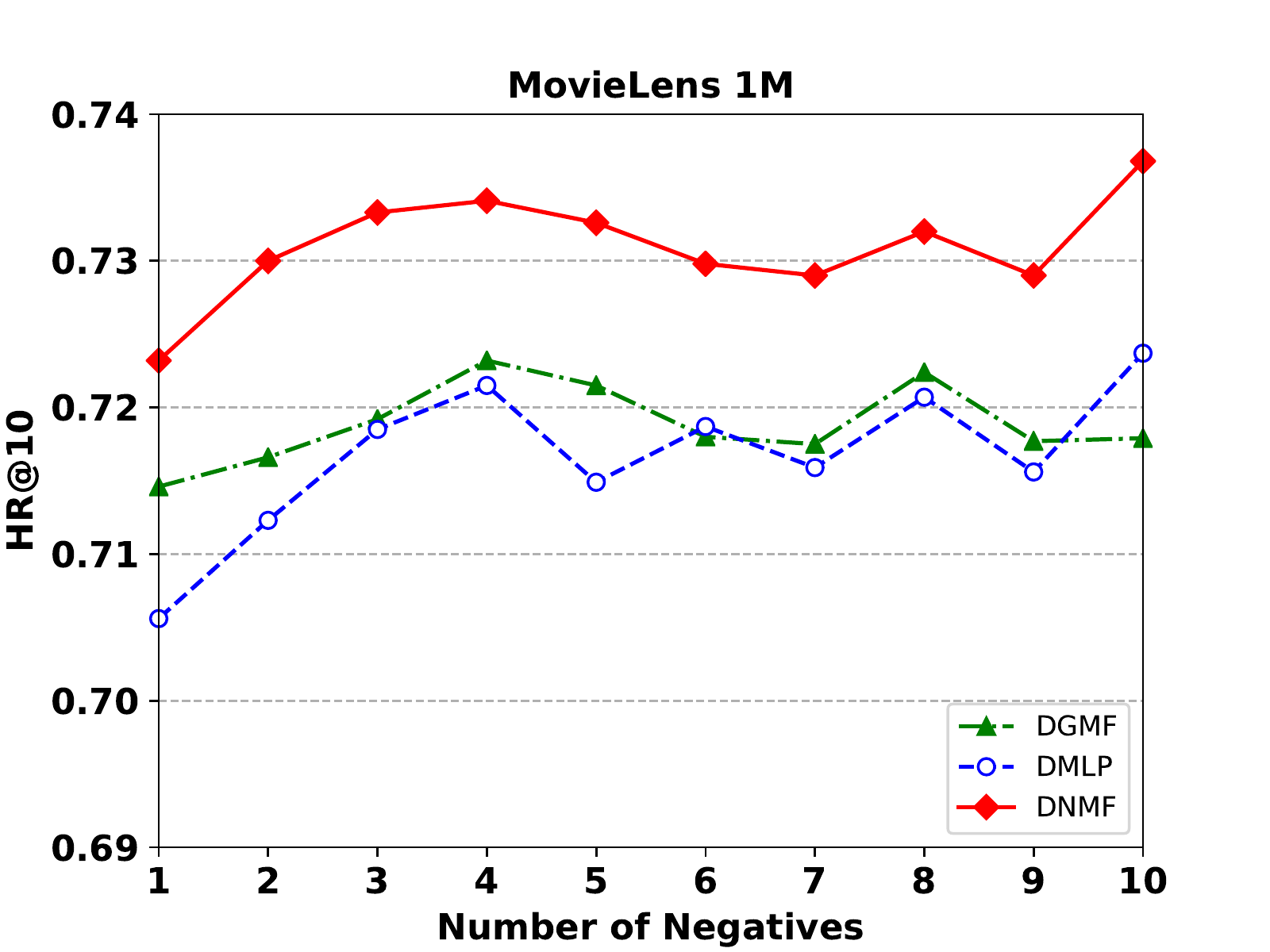}
    }
    \subfigure[MovieLens 1M --- NDCG@10]{
        \includegraphics[scale=0.25]{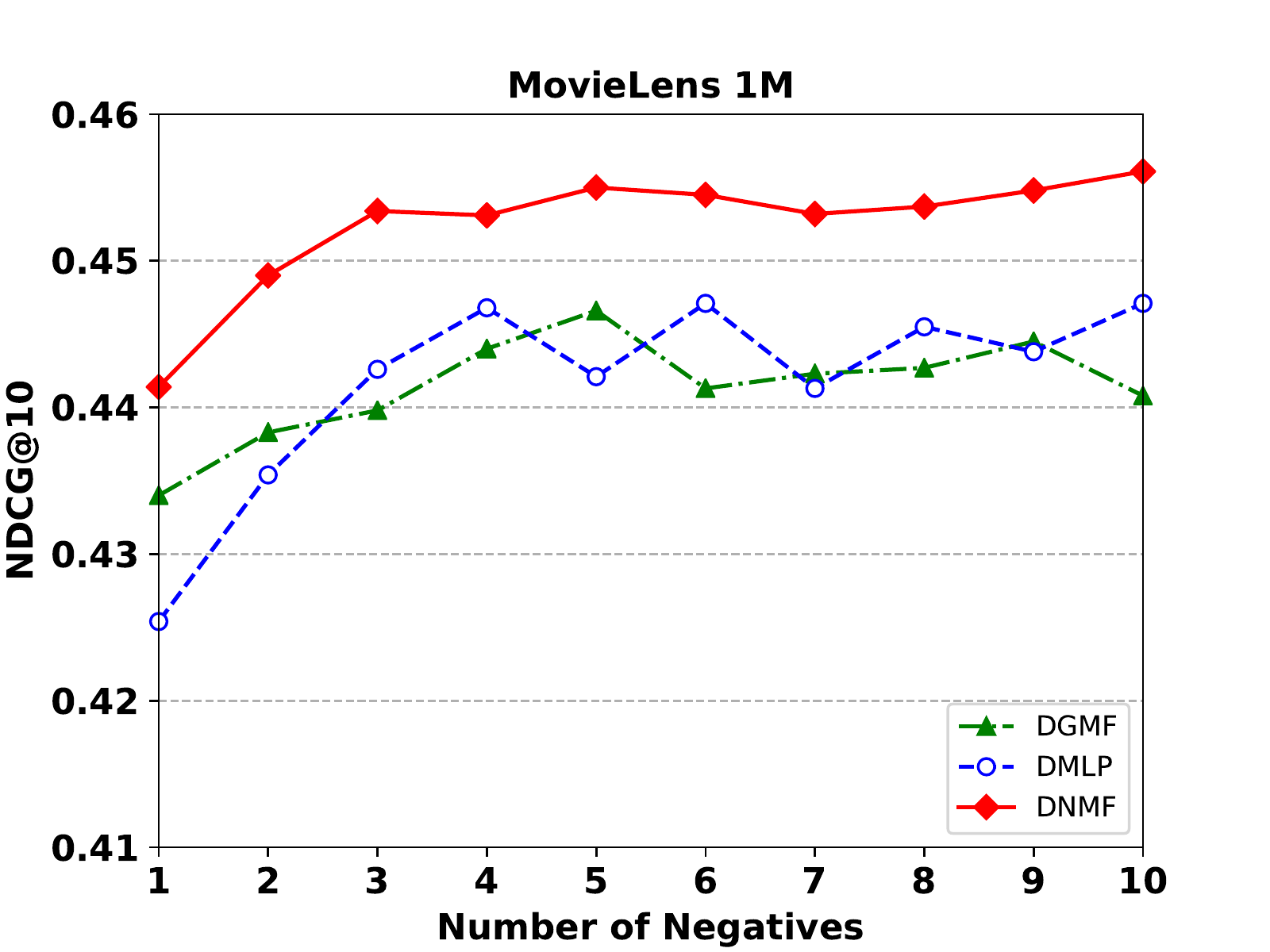}
    }
    \subfigure[Last.FM --- HR@10]{
        \includegraphics[scale=0.25]{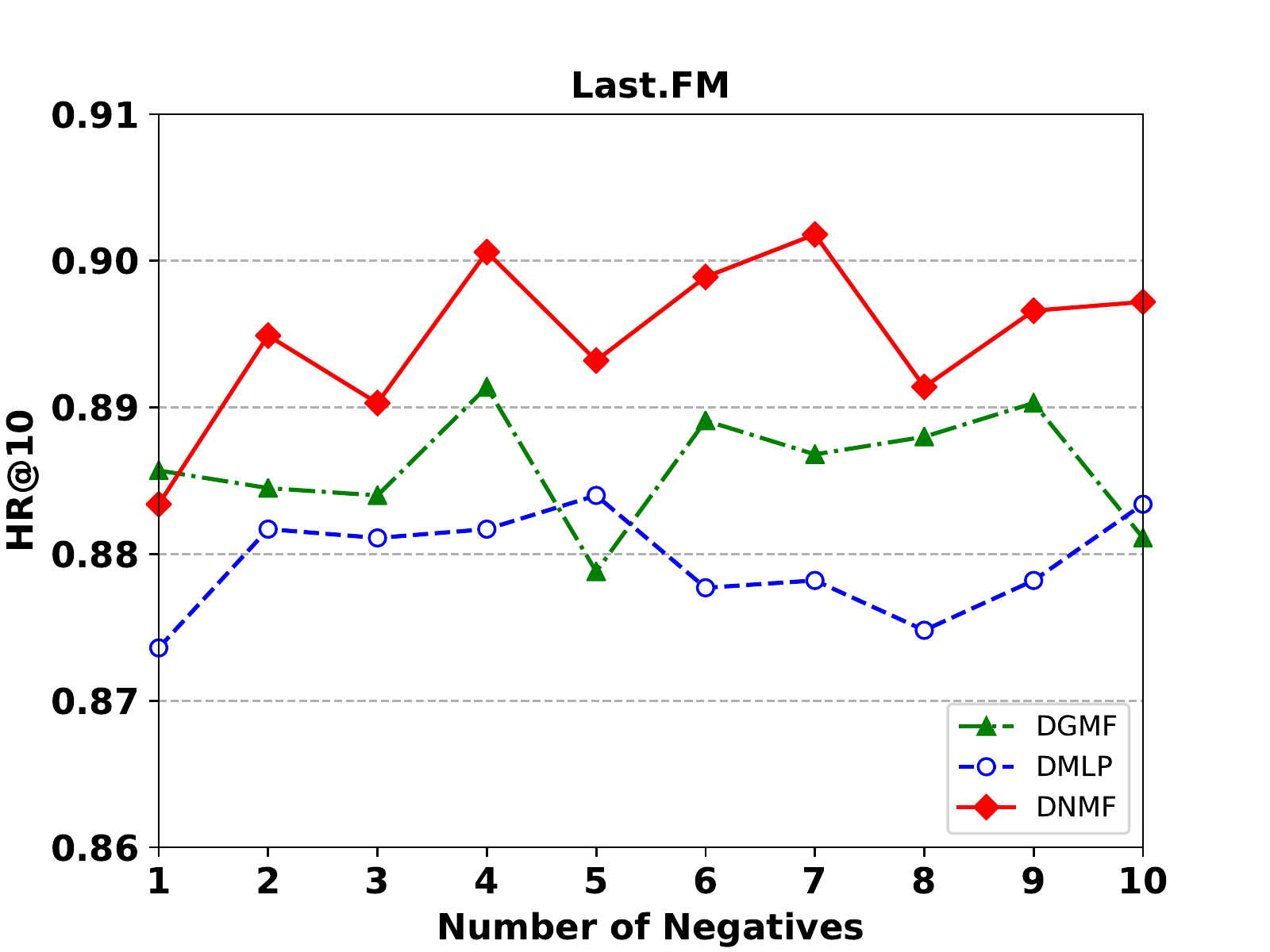}
    }
    \subfigure[Last.FM --- NDCG@10]{
        \includegraphics[scale=0.25]{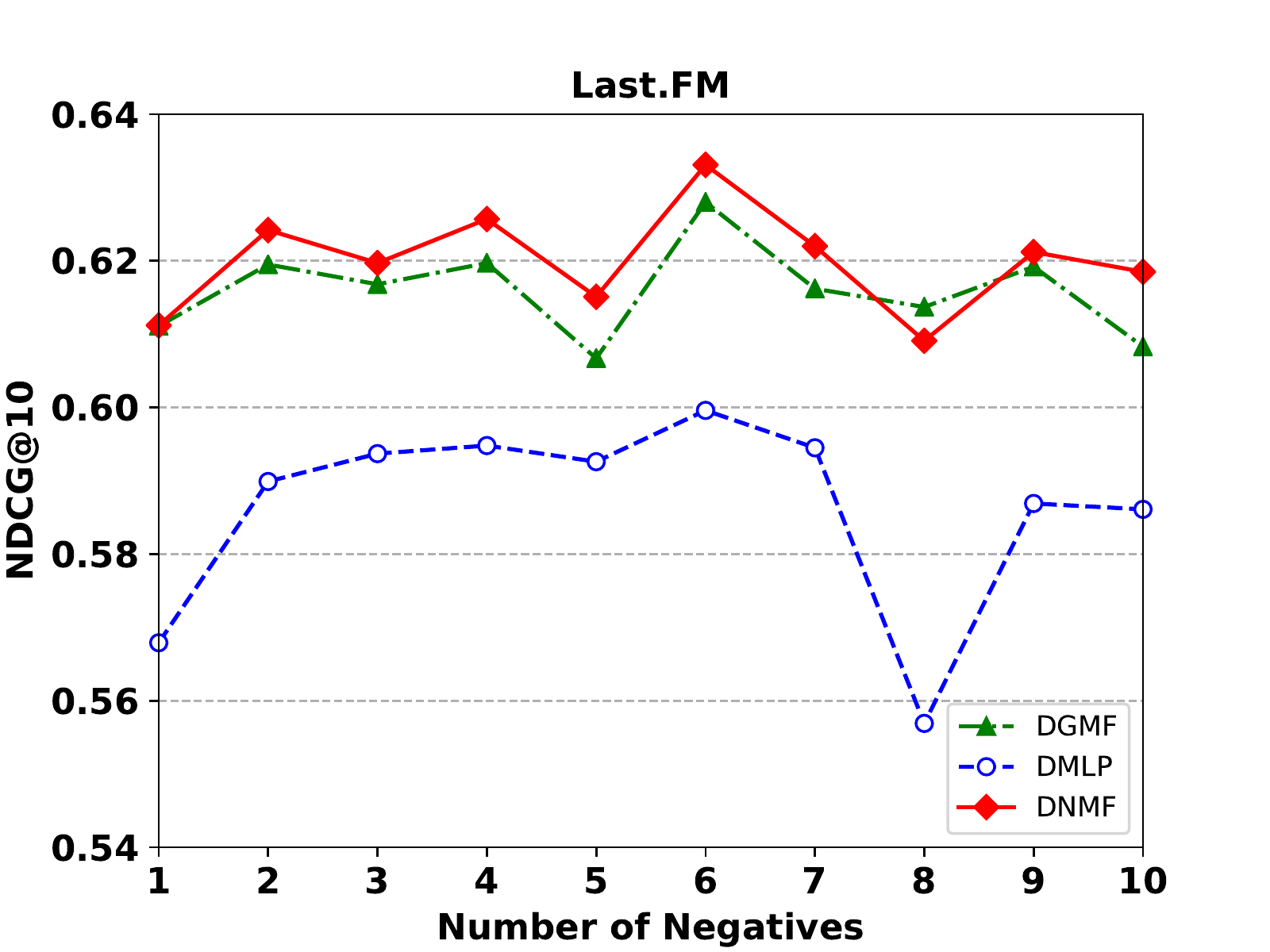}
    }
    \caption{Performance of DNCF methods w.r.t. the number of negative samples per positive instance (predictive factor = 64).}
    \label{fig:negative_ml-1m}
\end{figure*}

To analyse the impact of negative sampling for DNCF methods, we tested different negative sampling ratio, i.e. the number of negative samples per positive instance. Figure \ref{fig:negative_ml-1m} reports the performance of DNCF methods with respect to different negative sampling ratios on MovieLens 1M and Last.FM. As we can see, employing one negative instance is not enough and sampling more negative instances is beneficial to recommendation performance. For MovieLens 1M, the best performance is obtained when the negative sampling ratio is set to 10. 
For Last.FM, the best HR@10 is obtained when the negative sampling ratio is set to 7 while the best NDCG@10 is obtained when the negative sampling ratio is set to 6. 
To sum up, the optimal number of negative samples per positive instance is around 3 to 7, which is similar to the results shown in \cite{DBLP:conf/www/HeLZNHC17, DBLP:conf/aaai/DengHWLY19}. 
% Notice that when the sampling ratio is larger than 5, the performance of DNCF methods starts to drop on MovieLens 1M. It reveals that sampling more negative instances may degrade the performance.
Notice that it is not always a good idea to sampling more negative instances which requires not only more time to train the model but also more powerful machine with large memories to store the training data, even degrades the performance.

\subsection{Number of Hidden Layer in Network (RQ3)}

To test the effect of the number of hidden layers for DNCF approaches, we compared the performance of DMLP with respect to the number of hidden layers when the predictive factors is equal to 64 on four datasets.
The experimental results are provided in Table \ref{tab:hidden_layer}.
The Layer-3 denotes the DMLP method with three hidden layers, and similar notations for others.
We can find that to some extent stacking more non-linear hidden layers is beneficial to recommendation performance. This result demonstrates the effectiveness of using deep architecture for complex user-item interactions, which is in consistent with \cite{DBLP:conf/www/HeLZNHC17}.
% Furthermore, the performance of Layer-0 that has no hidden layers is as weak as non-personalized ItemPop, which is similar to \cite{DBLP:conf/www/HeLZNHC17}. This indicates that simple concatenation is insufficient to model user-item interactions.

\begin{figure*}[htbp]
    \centering
    \subfigure[MovieLens 1M --- HR@10]{
        \includegraphics[scale=0.25]{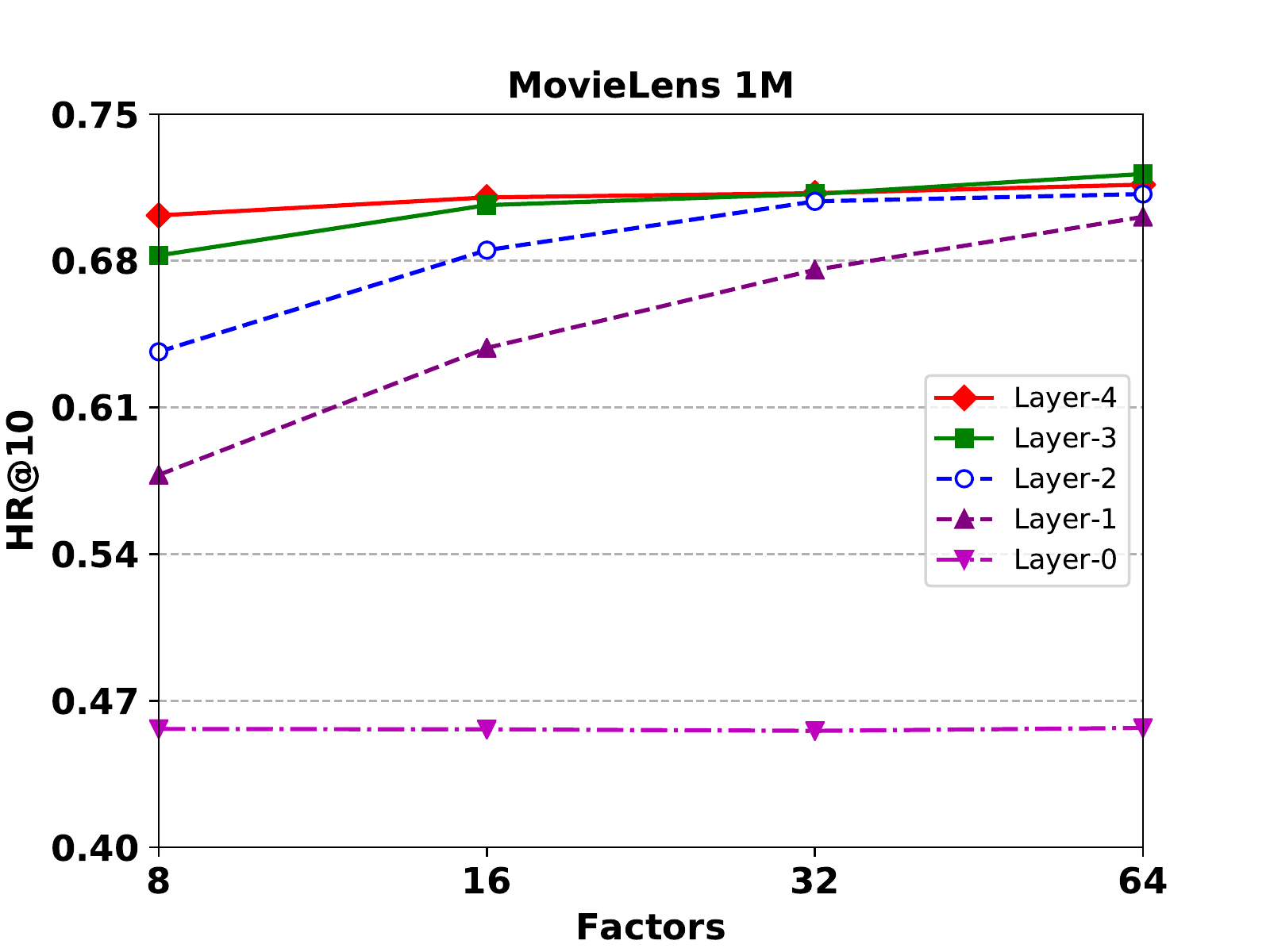}
    }
    \subfigure[MovieLens 1M --- NDCG@10]{
        \includegraphics[scale=0.25]{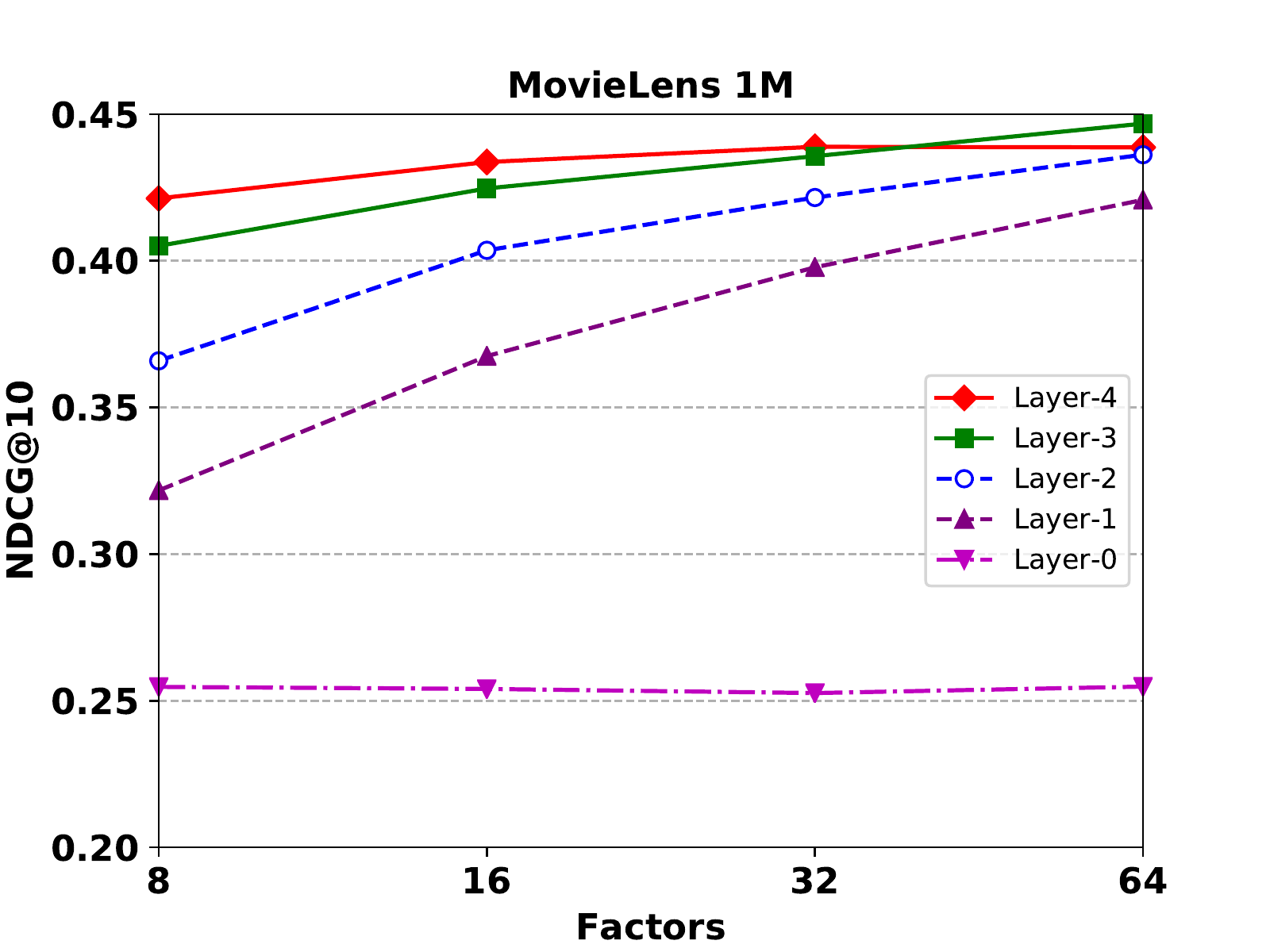}
    }
    \subfigure[Last.FM --- HR@10]{
        \includegraphics[scale=0.25]{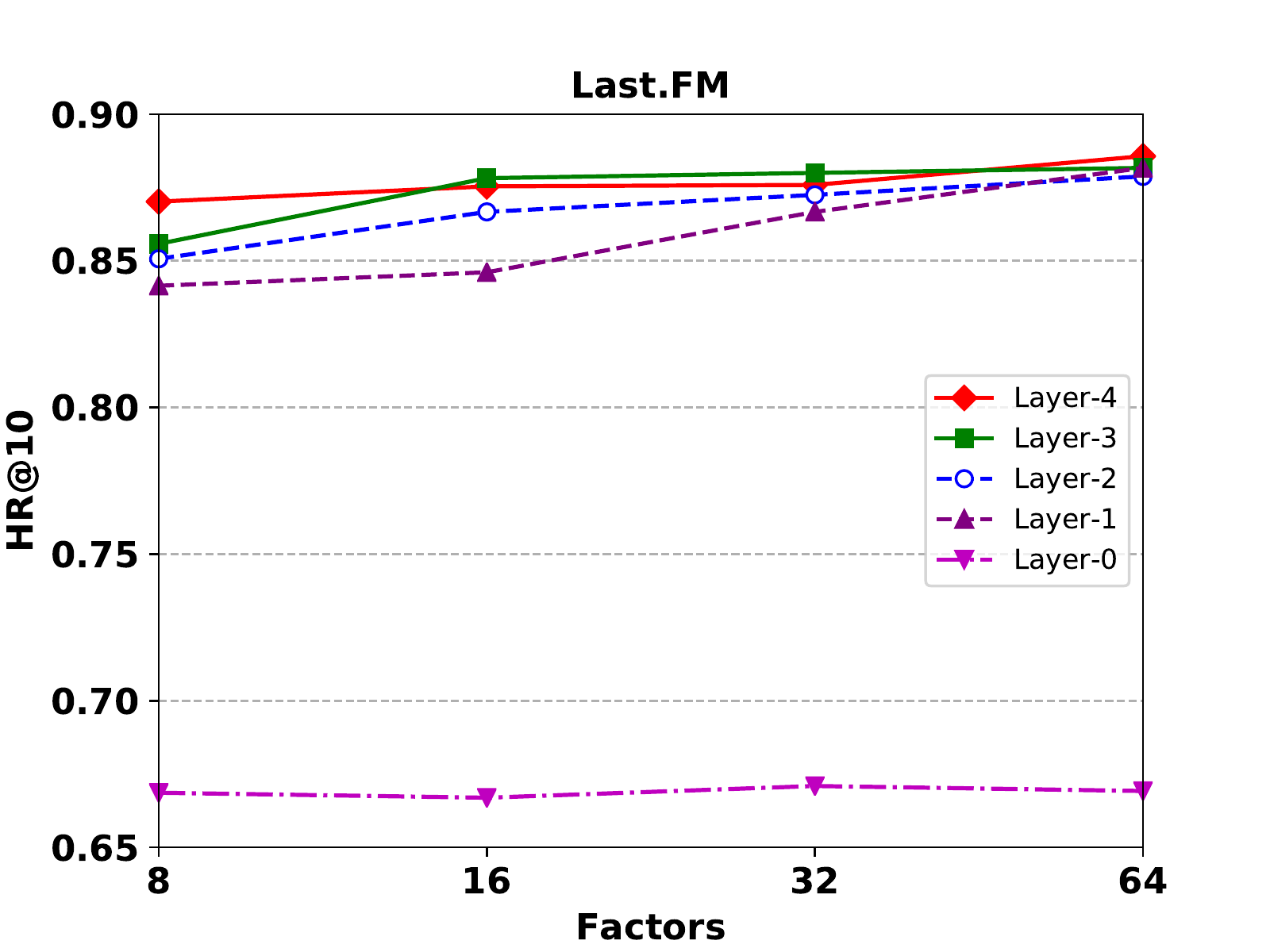}
    }
    \subfigure[Last.FM --- NDCG@10]{
        \includegraphics[scale=0.25]{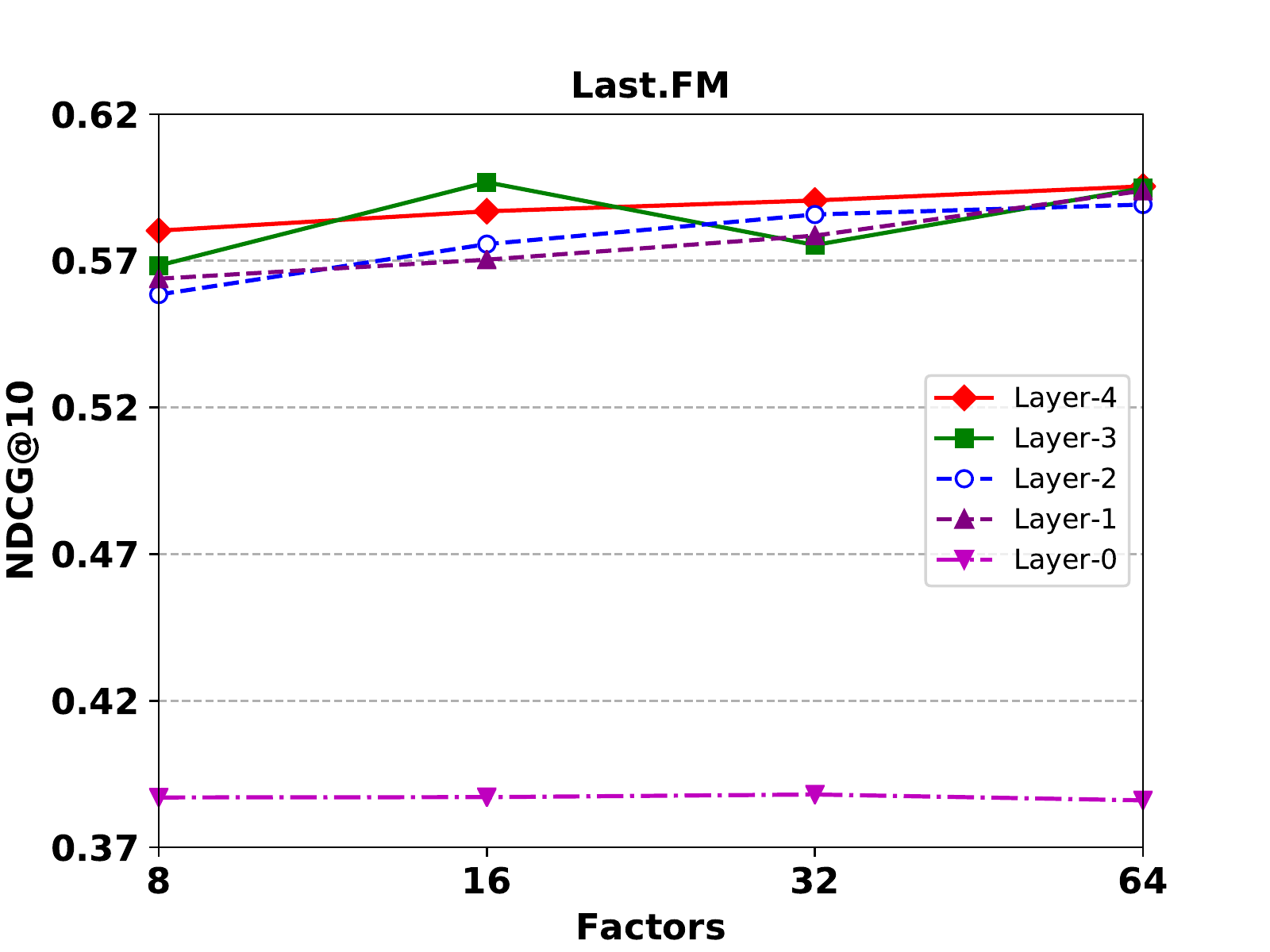}
    }
    \caption{Performance of DMLP with different layers on the MovieLens 1M and Last.FM.}
    \label{fig:hidden_layer}
\end{figure*}

To verify the above findings, we further investigated DMLP with different number of hidden layers and predictive factors on MovieLens 1M and Last.FM. The results are shown in Figure \ref{fig:hidden_layer}. We have the following observations: In most cases, stacking more layers yields better performance while the relative improvements decrease gradually with the increase of predictive factors.

\begin{table}[htbp]
    \caption{Performance of HR@10 and NDCG@10 with Different Layers at the Predictive Factor 64.}
    \label{tab:hidden_layer}
    \centering
    \resizebox{0.5\textwidth}{!}{
        \begin{tabular}{|c|c|c|c|c|c|}
            \hline
            \textbf{Datasets} &
            \textbf{Layer-0} & 
            \textbf{Layer-1} & 
            \textbf{Layer-2} & 
            \textbf{Layer-3} & 
            \textbf{Layer-4} \\
            \hline \hline
            
            \multicolumn{6}{|c|}{\textbf{HR@10}} \\
            \hline
            \textbf{MovieLens 1M} & 0.4570 & 0.7010 & 0.7119 & \textbf{0.7215} & 0.7164 \\
            \hline
            \textbf{Last.FM} & 0.6692 & 0.8817 & 0.8788 & 0.8817 & \textbf{0.8857}\\
            \hline
            \textbf{AMusic} & 0.2506 & 0.4200 & \textbf{0.4313} & 0.4257 & 0.4302\\
            \hline
            \textbf{AToy} & 0.2977 & 0.3787 & 0.3851 & \textbf{0.3985} & 0.3950\\
            \hline \hline
            
            \multicolumn{6}{|c|}{\textbf{NDCG@10}} \\
            \hline
            \textbf{MovieLens 1M} & 0.2548 & 0.4208 & 0.4362 & \textbf{0.4468} & 0.4387\\
            \hline
            \textbf{Last.FM} & 0.3860 & 0.5939 & 0.5892 & 0.5948 & \textbf{0.5954}\\
            \hline
            \textbf{AMusic} & 0.1301 & 0.2477 & 0.2547 & 0.2558 & \textbf{0.2596}\\
            \hline
            \textbf{AToy} & 0.1596 & 0.2201 & 0.2273 & \textbf{0.2353} & 0.2337\\
            \hline
        \end{tabular}
    }
\end{table}

\subsection{Effect of Embedding Combination Functions (RQ4)\label{section:dgmf_variants}}

\begin{figure*}[tbp]
    \centering
    \subfigure[MovieLens 1M --- HR@10]{
        \includegraphics[scale=0.25]{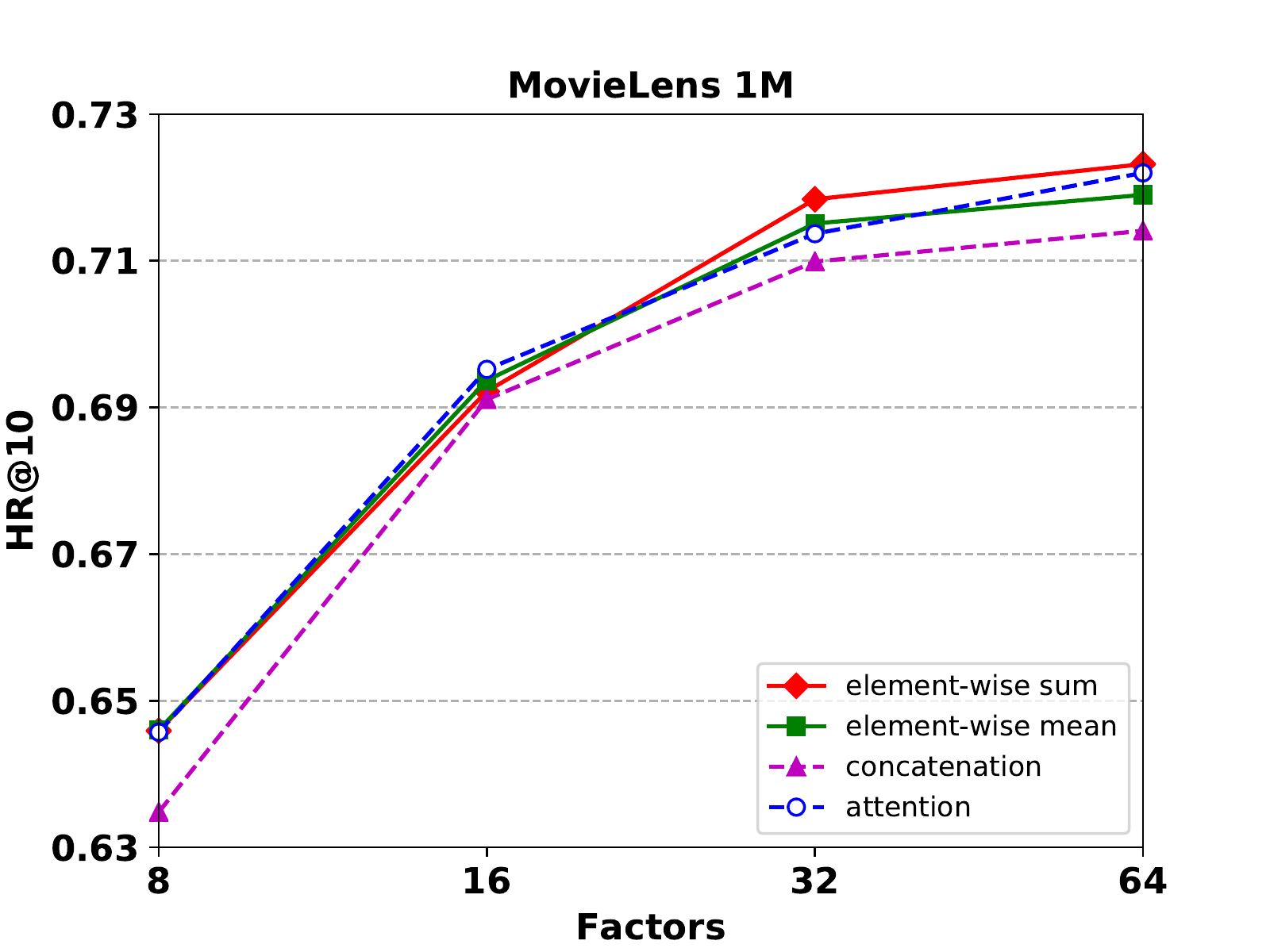}
    }
    \subfigure[MovieLens 1M --- NDCG@10]{
        \includegraphics[scale=0.25]{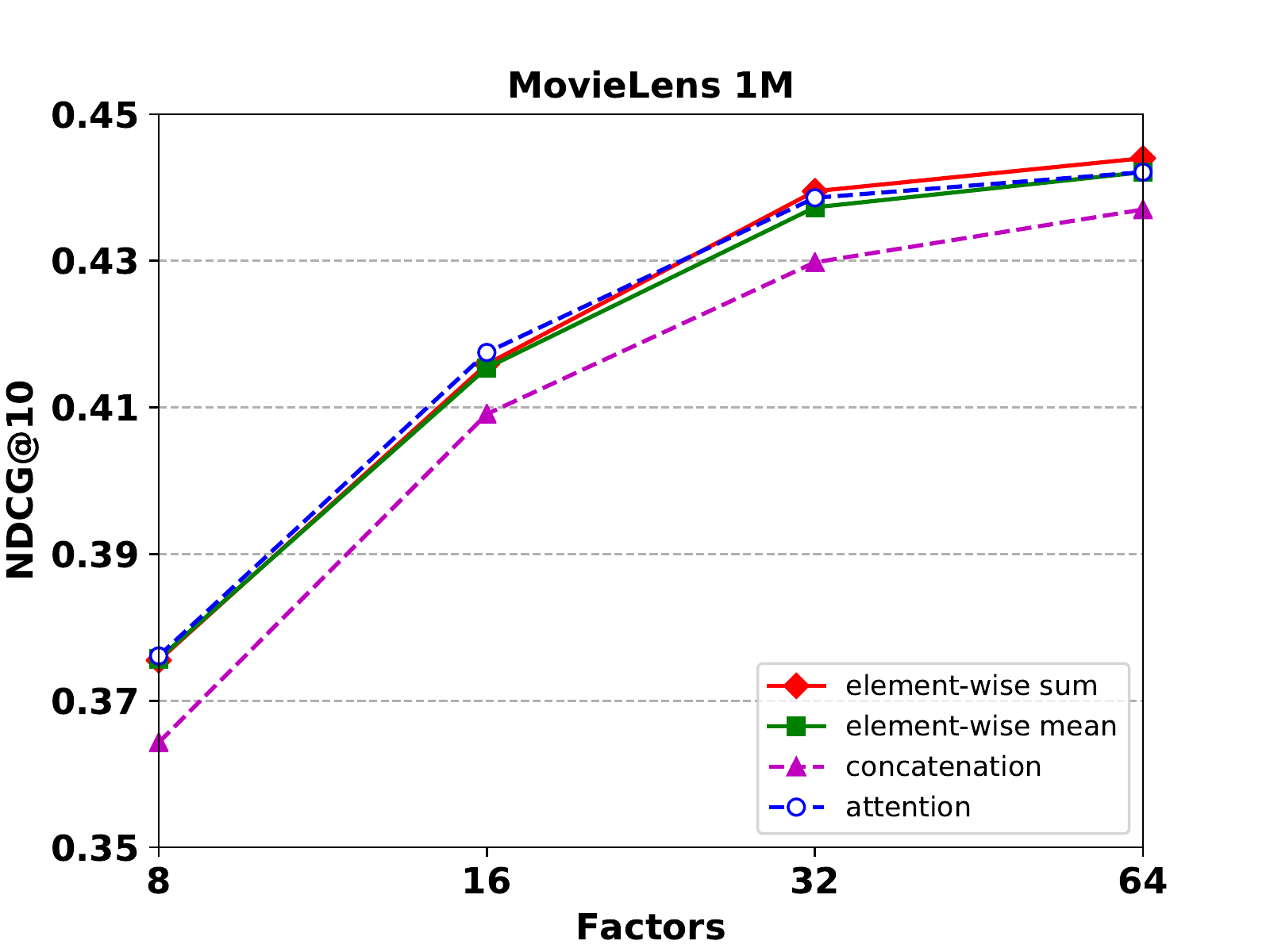}
    }
    \subfigure[Last.FM --- HR@10]{
        \includegraphics[scale=0.25]{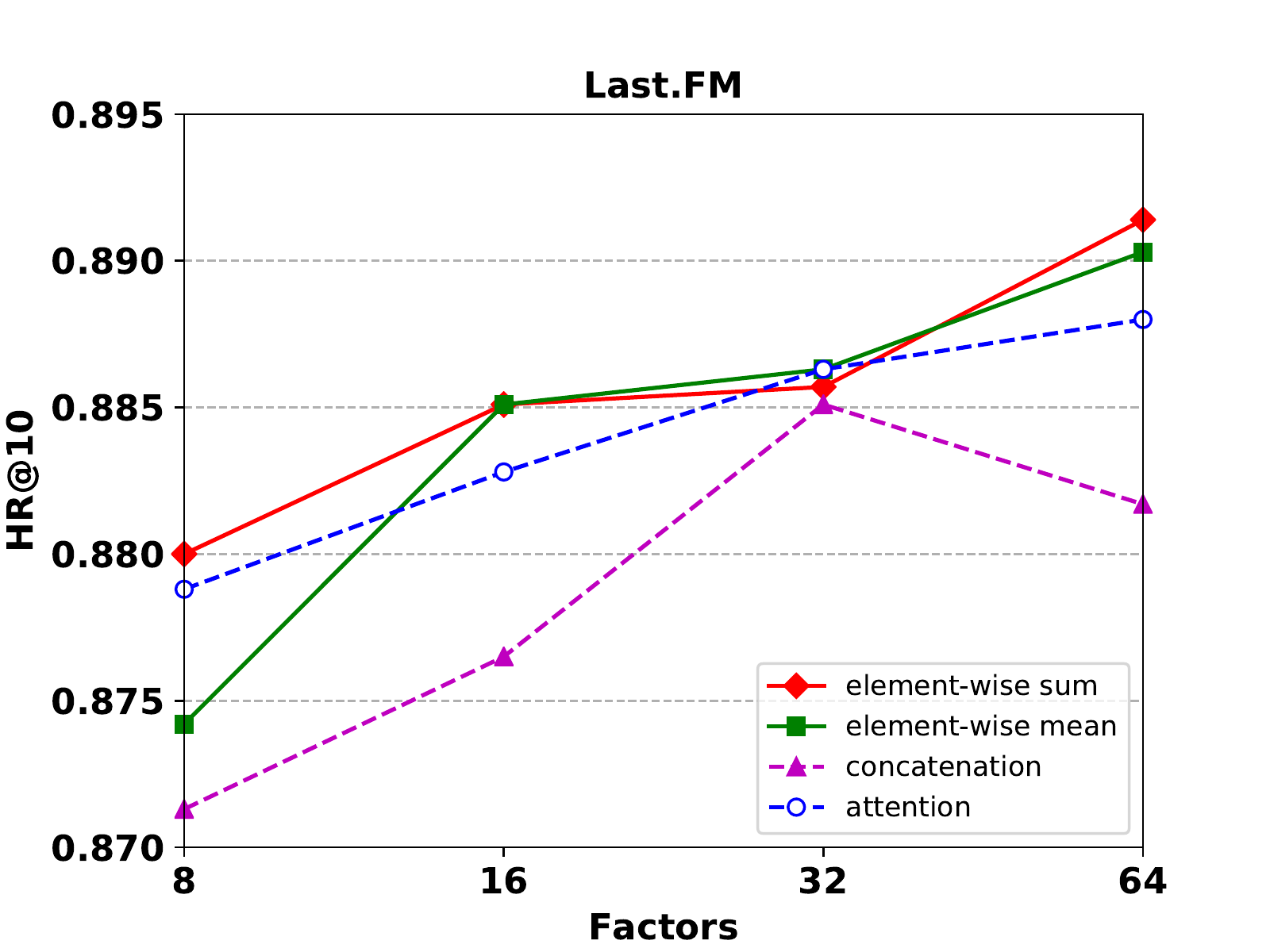}
    }
    \subfigure[Last.FM --- NDCG@10]{
        \includegraphics[scale=0.25]{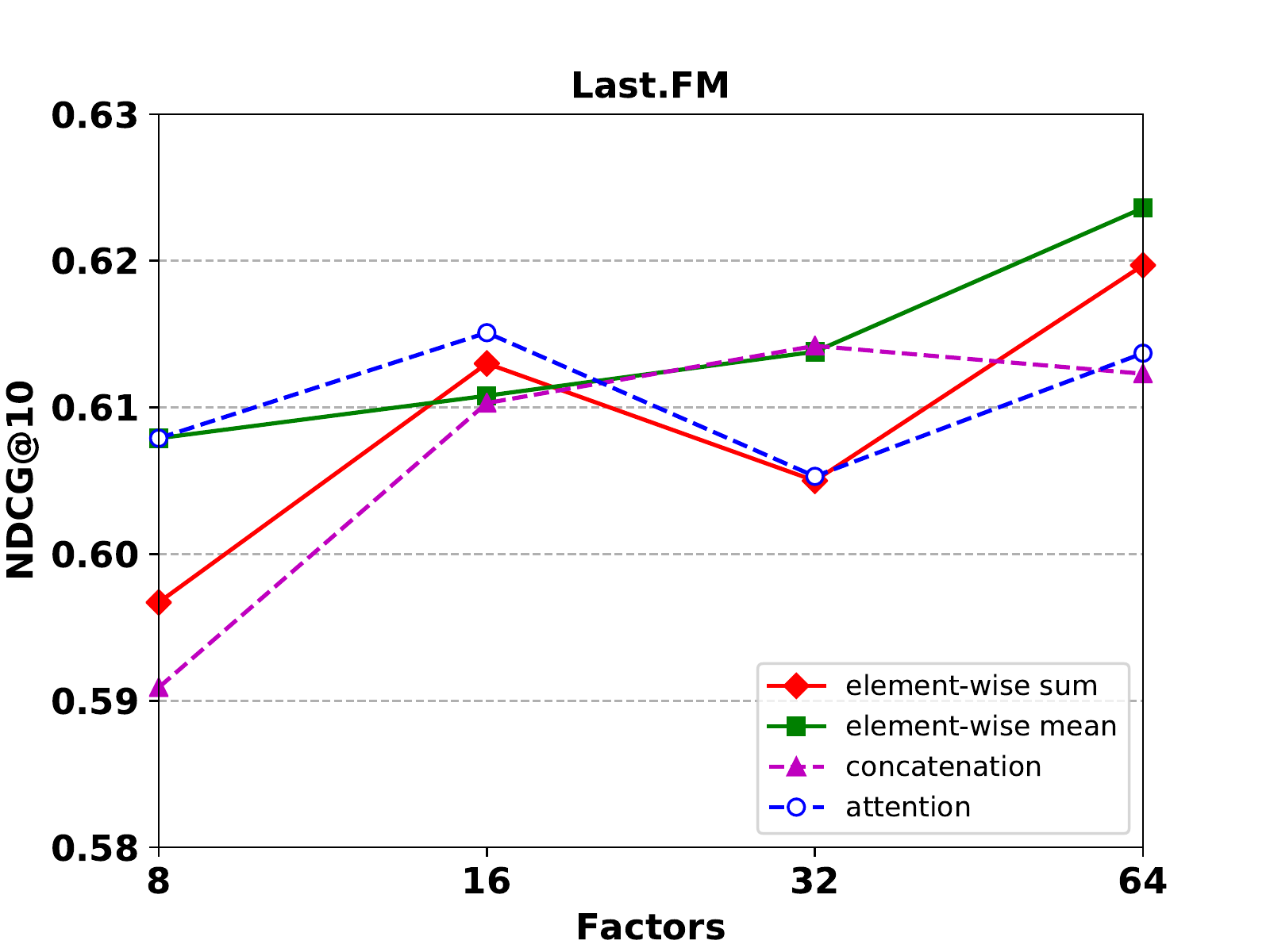}
    }
    \caption{Performance of different variants of DGMF on the MovieLens 1M and Last.FM.}
    \label{fig:dgmf_method}
\end{figure*}

We compared different methods to summary dual embeddings into one vector: \textit{element-wise sum, mean, concatenation} and \textit{attention}.
Table \ref{tab:dgmf_variants} shows the experimental results on four datasets when the predictive factors is set to 64.
We make the following observations: There is no one-size-fits-all method. Element-wise sum performs much better than other methods on MovieLens 1M. For Last.FM, element-wise sum also outperforms other methods in terms of HR while underperforms element-wise mean in terms of NDCG. However, concatenation achieves the best performance on both metrics on AMusic and element-wise mean outperforms other approaches on AToy. 
  
In order to verify the above conclusion, we further compared four methods with respect to different predictive factors on MovieLens 1M and Last.FM. The results are shown in Figure \ref{fig:dgmf_method}. It can be seen clearly that the optimal method varies with the increase of predictive factors on both datasets. In general, simple summation and mean perform as good as or even better than attention and concatenation. Moreover, they do not involve any additional trainable parameters. 

\begin{table}[tbp]
    \caption{Performance of Different Variants of DGMF at Predictive Factor 64.}
    \label{tab:dgmf_variants}
    \centering
    \resizebox{0.5\textwidth}{!}{
        \begin{tabular}{|c|c|c|c|c|}
             \hline
             \textbf{Methods} & 
             \textbf{MovieLens 1M} & 
             \textbf{Last.FM} & 
             \textbf{AMusic} & 
             \textbf{AToy} \\
             \hline \hline
             
             \multicolumn{5}{|c|}{\textbf{HR@10}} \\
             \hline
             \textbf{element-wise sum} & \textbf{0.7232} & \textbf{0.8914} & 0.4200 & 0.3994\\
             \hline
             \textbf{element-wise mean} & 0.7190 & 0.8903 & 0.4375 & \textbf{0.4198}\\
             \hline
             \textbf{concatenation} & 0.7141 & 0.8817 & \textbf{0.4516} & 0.4131\\
             \hline
             \textbf{attention} & 0.7220 & 0.8880 & 0.4127 & 0.4010\\
             \hline \hline
             
             \multicolumn{5}{|c|}{\textbf{NDCG@10}} \\
             \hline
             \textbf{element-wise sum} & \textbf{0.4440} & 0.6197 & 0.2584 & 0.2500\\
             \hline
             \textbf{element-wise mean} & 0.4421 & \textbf{0.6236} & 0.2744 & \textbf{0.2691}\\
             \hline
             \textbf{concatenation} & 0.4370 & 0.6123 & \textbf{0.2761} & 0.2669\\
             \hline
             \textbf{attention} & 0.4421 & 0.6137 & 0.2622 & 0.2555\\
             \hline
        \end{tabular}
    }
\end{table}

\section{Conclusion and Future Work\label{section:conclusion_futere_work}}
In this work, we explored dual-embedding based collaborative filtering methods for top-N recommendation. In addition to the primitive user and item embeddings, we obtained additional embedding for user and item based on their historical interactions from implicit feedback. In other words, we employed the items interacted by users to enhance user representation and useed the users once interacted with items to enrich item representation. Based on dual embeddings mentioned above, we devised a general framework DNCF and proposed three instantiations --- DMLP, DGMF and DNMF. We conducted comprehensive experiments on four real-world datasets and the corresponding experimental results demonstrated the superior performance of our proposed models compared with other state-of-the-art approaches for top-N item recommendation task.

In the future, we will study the following problems. First, all historical items (users) of a user (an item) contribute equally to the final history embedding in this work, which is an unrealistic assumption as reported in \cite{DBLP:journals/tkde/HeHSLJC18, DBLP:conf/ijcai/ChengSZH18}. So we would like to employ \textit{attention mechanism} \cite{DBLP:journals/corr/BahdanauCB14} to distinguish the importance of interacted items (users) when constructing historical interactions based embedding for each user (item). Second, auxiliary information can be used to further improve the representation of users and items, such as user reviews \cite{DBLP:conf/wsdm/ZhengNY17, DBLP:conf/kdd/LiuLDCG19}, item information \cite{DBLP:conf/kdd/WangWY15,DBLP:conf/kdd/LiS17}, knowledge base \cite{DBLP:conf/kdd/ZhangYLXM16, DBLP:conf/www/WangX000C20} and social networks \cite{DBLP:conf/aaai/GuoZY15, DBLP:conf/aaai/WangZYZ18}. Richer information usually leads to better performance. Third, we will also try to use different type of loss function, for example, BPR \cite{DBLP:conf/uai/RendleFGS09} to learn our models. Finally, Graph Convolutional Networks (GCNs) \cite{DBLP:conf/iclr/KipfW17} have attracted considerable research interest and some recent works \cite{DBLP:conf/kdd/YingHCEHL18, DBLP:conf/aaai/ChenWHZW20, DBLP:conf/sigir/0001DWLZ020} have employed GCNs to improve the performance of top-N recommendation. We're also very interested in exploring it to enhance the quality of the embedding.

\section*{Acknowledgment}

We would like to thank our anonymous reviewers for their helpful comments and valuable suggestions.

% references
% \bibliographystyle{ieeetr}
\bibliography{reference}

\end{document}